
\overfullrule=0pt
\newcount\mgnf\newcount\tipi\newcount\tipoformule\newcount\greco

\tipi=2          
\tipoformule=0   

\global\newcount\numsec\global\newcount\numfor
\global\newcount\numapp\global\newcount\numcap
\global\newcount\numfig\global\newcount\numpag
\global\newcount\numnf

\def\SIA #1,#2,#3 {\senondefinito{#1#2}%
\expandafter\xdef\csname #1#2\endcsname{#3}\else
\write16{???? ma #1,#2 e' gia' stato definito !!!!} \fi}

\def \FU(#1)#2{\SIA fu,#1,#2 }

\def\etichetta(#1){(\veroparagrafo.\veraformula)%
\SIA e,#1,(\veroparagrafo.\veraformula) %
\global\advance\numfor by 1%
\write15{\string\FU (#1){\equ(#1)}}%
\write16{ EQ #1 ==> \equ(#1)  }}
\def\etichettaa(#1){(A\veraappendice.\veraformula)
 \SIA e,#1,(A\veraappendice.\veraformula)
 \global\advance\numfor by 1
 \write15{\string\FU (#1){\equ(#1)}}
 \write16{ EQ #1 ==> \equ(#1) }}
\def\getichetta(#1){Fig. \verafigura
 \SIA g,#1,{\verafigura}
 \global\advance\numfig by 1
 \write15{\string\FU (#1){\graf(#1)}}
 \write16{ Fig. #1 ==> \graf(#1) }}
\def\retichetta(#1){\numpag=\pgn\SIA r,#1,{\verapagina}
 \write15{\string\FU (#1){\rif(#1)}}
 \write16{\rif(#1) ha simbolo  #1  }}
\def\etichettan(#1){(n\verocapitolo.\veranformula)
 \SIA e,#1,(n\verocapitolo.\veranformula)
 \global\advance\numnf by 1
\write16{\equ(#1) <= #1  }}

\newdimen\gwidth
\gdef\profonditastruttura{\dp\strutbox}
\def\senondefinito#1{\expandafter\ifx\csname#1\endcsname\relax}
\def\BOZZA{
\def\alato(##1){
 {\vtop to \profonditastruttura{\baselineskip
 \profonditastruttura\vss
 \rlap{\kern-\hsize\kern-1.2truecm{$\scriptstyle##1$}}}}}
\def\galato(##1){ \gwidth=\hsize \divide\gwidth by 2
 {\vtop to \profonditastruttura{\baselineskip
 \profonditastruttura\vss
 \rlap{\kern-\gwidth\kern-1.2truecm{$\scriptstyle##1$}}}}}
\def\verapagina{
{\romannumeral\number\numcap}.\number\numsec.\number\numpag}}

\def\alato(#1){}
\def\galato(#1){}
\def\veroparagrafo{\number\numsec}\def\veraformula{\number\numfor}
\def\veraappendice{\number\numapp}
\def\verapagina{\number\pageno}\def\veranformula{\number\numnf}
\def\verafigura{{\romannumeral\number\numcap}.\number\numfig}
\def\verocapitolo{\number\numcap}\def\veranformula{\number\numnf}
\def\Eqn(#1){\eqno{\etichettan(#1)\alato(#1)}}
\def\eqn(#1){\etichettan(#1)\alato(#1)}

\def\Eq(#1){\eqno{\etichetta(#1)\alato(#1)}}
\def\eq(#1){\etichetta(#1)\alato(#1)}
\def\Eqa(#1){\eqno{\etichettaa(#1)\alato(#1)}}
\def\eqa(#1){\etichettaa(#1)\alato(#1)}
\def\dgraf(#1){\getichetta(#1)\galato(#1)}
\def\drif(#1){\retichetta(#1)}

\def\eqv(#1){\senondefinito{fu#1}$\clubsuit$#1\else\csname fu#1\endcsname\fi}
\def\equ(#1){\senondefinito{e#1}\eqv(#1)\else\csname e#1\endcsname\fi}
\def\graf(#1){\senondefinito{g#1}\eqv(#1)\else\csname g#1\endcsname\fi}
\def\rif(#1){\senondefinito{r#1}\eqv(#1)\else\csname r#1\endcsname\fi}
\def\bib[#1]{[#1]\numpag=\pgn
\write13{\string[#1],\verapagina}}

\def\include#1{
\openin13=#1.aux \ifeof13 \relax \else
\input #1.aux \closein13 \fi}

\openin14=\jobname.aux \ifeof14 \relax \else
\input \jobname.aux \closein14 \fi
\openout15=\jobname.aux
\openout13=\jobname.bib


\ifnum\tipoformule=1\let\Eq=\eqno\def\eq{}\let\Eqa=\eqno\def\eqa{}
\def\equ{}\fi


{\count255=\time\divide\count255 by 60 \xdef\hourmin{\number\count255}
        \multiply\count255 by-60\advance\count255 by\time
   \xdef\hourmin{\hourmin:\ifnum\count255<10 0\fi\the\count255}}

\def\oramin{\hourmin }

\def\data{\number\day/\ifcase\month\or january \or february \or march \or
april \or may \or june \or july \or august \or september
\or october \or november \or december \fi/\number\year;\ \oramin}

\setbox200\hbox{$\scriptscriptstyle \data $}

\newcount\pgn \pgn=1
\def\foglio{\number\numsec:\number\pgn
\global\advance\pgn by 1}
\def\foglioa{A\number\numsec:\number\pgn
\global\advance\pgn by 1}

\footline={\hss\tenrm\folio\hss}
\nopagenumbers
\def\TIPIO{
\font\setterm=amr7 
\def \settepunti{\def\rm{\fam0\setterm}
\textfont0=\setterm   
\normalbaselineskip=9pt\normalbaselines\rm
}\let\nota=\settepunti}

\def\TIPITOT{
\font\twelverm=cmr12
\font\twelvei=cmmi12
\font\twelvesy=cmsy10 scaled\magstep1
\font\twelveex=cmex10 scaled\magstep1
\font\twelveit=cmti12
\font\twelvett=cmtt12
\font\twelvebf=cmbx12
\font\twelvesl=cmsl12
\font\ninerm=cmr9
\font\ninesy=cmsy9
\font\eightrm=cmr8
\font\eighti=cmmi8
\font\eightsy=cmsy8
\font\eightbf=cmbx8
\font\eighttt=cmtt8
\font\eightsl=cmsl8
\font\eightit=cmti8
\font\sixrm=cmr6
\font\sixbf=cmbx6
\font\sixi=cmmi6
\font\sixsy=cmsy6
\font\twelvetruecmr=cmr10 scaled\magstep1
\font\twelvetruecmsy=cmsy10 scaled\magstep1
\font\tentruecmr=cmr10
\font\tentruecmsy=cmsy10
\font\eighttruecmr=cmr8
\font\eighttruecmsy=cmsy8
\font\seventruecmr=cmr7
\font\seventruecmsy=cmsy7
\font\sixtruecmr=cmr6
\font\sixtruecmsy=cmsy6
\font\fivetruecmr=cmr5
\font\fivetruecmsy=cmsy5
\textfont\truecmr=\tentruecmr
\scriptfont\truecmr=\seventruecmr
\scriptscriptfont\truecmr=\fivetruecmr
\textfont\truecmsy=\tentruecmsy
\scriptfont\truecmsy=\seventruecmsy
\scriptscriptfont\truecmr=\fivetruecmr
\scriptscriptfont\truecmsy=\fivetruecmsy
\def \eightpoint{\def\rm{\fam0\eightrm}
\textfont0=\eightrm \scriptfont0=\sixrm \scriptscriptfont0=\fiverm
\textfont1=\eighti \scriptfont1=\sixi   \scriptscriptfont1=\fivei
\textfont2=\eightsy \scriptfont2=\sixsy   \scriptscriptfont2=\fivesy
\textfont3=\tenex \scriptfont3=\tenex   \scriptscriptfont3=\tenex
\textfont\itfam=\eightit  \def\it{\fam\itfam\eightit}%
\textfont\slfam=\eightsl  \def\sl{\fam\slfam\eightsl}%
\textfont\ttfam=\eighttt  \def\tt{\fam\ttfam\eighttt}%
\textfont\bffam=\eightbf  \scriptfont\bffam=\sixbf
\scriptscriptfont\bffam=\fivebf  \def\bf{\fam\bffam\eightbf}%
\tt \ttglue=.5em plus.25em minus.15em
\setbox\strutbox=\hbox{\vrule height7pt depth2pt width0pt}%
\normalbaselineskip=9pt
\let\sc=\sixrm  \let\big=\eightbig  \normalbaselines\rm
\textfont\truecmr=\eighttruecmr
\scriptfont\truecmr=\sixtruecmr
\scriptscriptfont\truecmr=\fivetruecmr
\textfont\truecmsy=\eighttruecmsy
\scriptfont\truecmsy=\sixtruecmsy
}\let\nota=\eightpoint}

\newfam\msbfam   
\newfam\truecmr  
\newfam\truecmsy 
\newskip\ttglue
\ifnum\tipi=0\TIPIO \else\ifnum\tipi=1 \TIPI\else \TIPITOT\fi\fi

\global\newcount\numpunt

\magnification=\magstephalf
\baselineskip=16pt
\parskip=8pt

\def\a{\alpha}
\def\b{\beta}
\def\d{\delta}
\def\e{\epsilon}

\def\g{\gamma}
\def\k{\kappa}
\def\l{\lambda}

\def\s{\sigma}
\def\t{\tau}

\def\o{\omega}

\def\G{\Gamma}
\def\O{\Omega}

\def\Th{\Theta}

\def\E{{I\kern-.25em{E}}}
\def\N{{I\kern-.22em{N}}}
\def\M{{I\kern-.22em{M}}}
\def\R{{I\kern-.22em{R}}}
\def\Z{{Z\kern-.5em{Z}}}
\def\1{{1\kern-.25em\hbox{\rm I}}}
\def\eu{{1\kern-.25em\hbox{\sm I}}}

\def\C{{C\kern-.75em{C}}}
\def\P{{I\kern-.25em{P}}}



\def\AA{{\cal A}}
\def\BB{{\cal B}}

\def\FF{{\cal F}}

\def\KK{{\cal K}}
\def\SS{{\cal S}}

\def\WW{{\cal W}}

\def\QQ{{\cal Q}}
\def\A{{\cal A}}

\def\chap #1#2{\line{\ch #1\hfill}\numsec=#2\numfor=1}

\def\ba{{\backslash}}

\def\sminn{\hbox{\ftn int}\,}
\def\smcl{\hbox{\ftn cl}\,}


\newcount\foot
\foot=1
\def\note#1{\footnote{${}^{\number\foot}$}{\ftn #1}\advance\foot by 1}
\def\tag #1{\eqno{\hbox{\rm(#1)}}}
\def\frac#1#2{{#1\over #2}}

\def\text#1{\quad{\hbox{#1}}\quad}
\def\newpage{\vfill\eject}
\def\proposition #1{\noindent{\thbf Proposition #1:}}

\def\theo #1{\noindent{\thbf Theorem #1: }}
\def\lemma #1{\noindent{\thbf Lemma #1: }}

\def\corollary #1{\noindent{\thbf Corollary #1: }}
\def\proof{{\noindent\pr Proof: }}

\def\endproof{$\diamondsuit$}
\def\remark{\noindent{\bf Remark: }}
\def\thanks{\noindent{\bf Aknowledgements: }}
\font\pr=cmbxsl10
\font\thbf=cmbxsl10 scaled\magstephalf

\font\ch=cmbx12
\font\ftn=cmr8

\font\it=cmti10
\font\bf=cmbx10
\font\sm=cmr7


\font\tit=cmbx12
\font\aut=cmbx12
\font\aff=cmsl12
\def\s{\char'31}
\nopagenumbers
{$  $}
\vskip2truecm
\centerline{\tit AN ALMOST SURE LARGE DEVIATION PRINCIPLE}
\vskip.2truecm
\centerline{\tit FOR THE HOPFIELD MODEL\footnote{${}^\#$}{\ftn Work
partially supported by the Commission of the European Communities
under contract  CHRX-CT93-0411}}
 \vskip1cm
\centerline{\aut Anton Bovier
\footnote{${}^1$}{\ftn e-mail:
bovier@iaas-berlin.d400.de}}
\vskip.1truecm
\centerline{\aff Weierstra\s {}--Institut}
\centerline{\aff f\"ur Angewandte Analysis und Stochastik}
\centerline{\aff Mohrenstrasse 39, D-10117 Berlin, Germany}
\vskip.5truecm
\centerline{\aut  V\'eronique Gayrard\footnote{${}^2$}{\ftn
e-mail: gayrard@cpt.univ-mrs.fr}}
\vskip.1truecm
\centerline{\aff Centre de Physique Th\'eorique - CNRS}
\centerline{\aff Luminy, Case 907}
\centerline{\aff F-13288 Marseille Cedex 9, France}
\vskip3truecm\rm
\def\s{\sigma}
\noindent {\bf Abstract:}
We prove a large deviation principle
for the finite dimensional marginals of the Gibbs distribution
of the macroscopic `overlap'-parameters in the Hopfield model
in the case where the number of random `patterns', $M$, as a
function of the system size $N$ satisfies $\limsup  M(N)/N=0$.
In this case the rate function is independent of the disorder for almost
all realizations of the patterns.


$ {} $
\newpage

\chap{1. Introduction}1

  Mean field models in statistical mechanics furnish nice examples
for the interpretation of thermodynamics as the theory of large deviation
for Gibbs measures of microscopically defined statistical mechanics systems
[E].
Roughly speaking, in such models the Hamiltonian is only a function of
(extensive) `macroscopic' quantities (density, magnetization,etc.)
of the system. In the thermodynamic limit,
the distribution of these quantities is expected to be concentrated on
a sharp value and to satisfy a large deviation principle. The corresponding
rate functions are then the thermodynamic potentials (free energy, pressure)
that govern the macroscopic response of the system to external
(intensive) conditions.
The classical paradigm of the theory is that the number of   relevant
macroscopic variables  is excessively small (order of 10)
compared to the number of
microscopic variables (order of $10^{23}$)   .

   Over the last decade, the formalism of statistical mechanics and
thermodynamics has found increasing applications in systems
in which the macroscopic behaviour is far more complex and described by
a `large' number of variables. Such systems can be found in biology
(heteropolymers, neural networks) but also in the domain of disordered
solids, and in particular spin glasses. Some fundamental aspects of
these ideas are discussed in an interesting recent paper by Parisi [P].
For such systems,
many basic problems are not very well understood, and many technical aspects
defy a mathematical investigation at the present time. An interesting
toy model (that nonetheless has also  practical relevance)
where this situation can be studied and for which
mathematical results are available, is the Hopfield model
[FP1,Ho]. This model
is a mean field spin system in the sense explained  above. However,
the Hamiltonian, instead of being a function of few macroscopic variables
is a function of macroscopic variables that are random functions of the
microscopic ones and those number tends to infinity with the size of the
system in a controllable way. More specifically, the model is
defined as follows.

Let $\SS_N\equiv \{-1,1\}^N$ denote the set  of functions
$\s:\{1,\dots,N\}\rightarrow \{-1,1\}$, and set $\SS\equiv \{-1,1\}^{\N}$.
We call $\s$ a spin configuration and denote by $\s_i$ the value of $\s$
at $i$. Let $(\O,\FF,\P)$ be an abstract probability space and let
$\xi^\mu_i$, $i,\mu\in \N$, denote a family of independent
identically distributed  random variables
on this space. For the purposes of this paper we will assume that
$\P[\xi_i^\mu=\pm 1]=\frac 12$, but more general distributions
can be considered.
We will write $\xi^\mu[\o]$ for the $N$-dimensional random
vector whose $i$-th
component is given by $\xi_i^\mu[\o]$
and call such a vector a `pattern'. On the other hand, we use the
notation
$\xi_i[\o]$ for the $M$-dimensional vector with the same components.
When we write $\xi[\o]$ without indices, we frequently will consider
it as an $M\times N$ matrix and we write $\xi^t[\o]$ for the transpose
of this matrix. Thus, $\xi^t[\o]\xi[\o]$ is the $M\times M$ matrix
whose elements are $\sum_{i=1}^N\xi_i^\mu[\o]\xi_i^\nu[\o]$. With this
in mind we will use throughout the paper a vector notation with
$(\cdot,\cdot)$ standing for the scalar product in whatever space the
argument may lie. E.g. the expression $(y,\xi_i)$ stands for
$\sum_{\mu=1}^M\xi_i^\mu y_\mu$, etc.

We define random maps $m_N^\mu[\o]:\SS_N\rightarrow [-1,1]$
through\note{We will make the dependence of random quantities on the
             random parameter $\o$ explicit by an added $[\o]$ whenever we want
             to stress it. Otherwise, we will frequently drop the
             reference to $\o$ to simplify the notation.}
$$
m_N^\mu[\o](\s)\equiv \frac 1N\sum_{i=1}^N \xi_i^\mu[\o]\s_i
\Eq(1.1)
$$
Naturally, these maps `compare' the configuration $\s$ globally to the
random configuration $\xi^\mu[\o]$. A Hamiltonian is now defined as the
simplest negative function of these variables, namely
$$
H_N[\o](\s)\equiv -\frac N2\sum_{\mu=1}^{M(N)} \left(m_N^\mu[\o](\s)\right)^2
\Eq(1.2)
$$
where $M(N)$ is some, generally increasing, function that crucially influences
the properties of the model. With $\|\cdot\|_2$ denoting the
$\ell_2$-norm in $\R^M$,  \eqv(1.2) can be written in  the compact form
$$
H_N[\o](\s)= -\frac N2 \left\|m_N[\o](\s)\right\|_2^2
\Eq(1.3)
$$
Also, $(\cdot , \cdot)$ will stand throughout for the scaler product of the
two argument in whatever space they may lie in.

Through this Hamiltonian we define in a natural way finite volume
Gibbs measures on $\SS_N$ via
$$
\mu_{N,\b}[\o](\s)\equiv \frac 1{Z_{N,\b}[\o]}e^{-\b H_N[\o](\s)}
\Eq(1.4)
$$
and the induced distribution of the overlap parameters
$$
\QQ_{N,\b}[\o]\equiv \mu_{N,\b}[\o]\circ m_N[\o]^{-1}
\Eq(1.5)
$$
The normalizing factor $Z_{N,\b}[\o]$, given by
$$
Z_{N,\b}[\o]\equiv 2^{-N}\sum_{\s\in \SS_N}e^{-\b H_N[\o](\s)}
\equiv \E_\s e^{-\b H_N[\o](\s)}
\Eq(1.5bis)
$$
is called the partition function.

This model has been studied very heavily in the physics literature. As a basic
introduction to what is commonly believed about its properties, we refer to the
seminal paper by Amit, Gutfreund and Sompolinsky [AGS].
Over the last few years, a considerable amount of mathematically rigorous
results on these measures has been established
[BG1,BGP1,BGP2,BGP3,K,N,KP,KPa,ST,PST].
It is known that under the hypothesis that $\limsup_{N\uparrow\infty}
M(N)/N=0$ weak limits can be constructed for which the
$\QQ_N$ converge to Dirac measures in $\R^\infty$ [BGP1]. Disjoint weak
limits have also been constructed in the case where
 $\limsup_{N\uparrow\infty}
M(N)/N=\a>0$, for small $\a$ in [BGP3].
In this note we restrict our attention
to the case $\a=0$ and the question to what extent a large deviation
principle (LDP) for the distribution of the macroscopic overlaps can be proven.

A first step in this direction had been taken already in [BGP2]. There, a
LDP was proven, but only under the restrictive assumption
$M(N)<\frac {\ln N}{\ln 2}$, while only a weaker result concerning the
existence of the convex hull of the rate function was proven in the general
case $\a=0$ in a rather indirect way. The first LDP in the Hopfield
model was proven earlier by Comets [Co] for the case of a finite
number of patterns.
Here we prove a LDP under more natural, and probably optimal, assumptions.

Since the overlap parameters form a vector in a space of unbounded
dimension, the most natural setting for a LDP is to consider
the finite dimensional marginals. Let $I\subset \N$ be a finite set of integers
and let $\R^I\subset \R^{\N}$ denote the corresponding subspace and finally
let $\Pi_I$ denote the canonical projection from $\R^p$ onto $\R^I$
for all $p$ for which $I\subset\{1,\dots,p\}$.
Without loss of generality we can and will assume in the sequel that
$I=\{1,\dots,|I|\}$.
Let us introduce the maps $n_{|I|}:[-1,1]^{2^p}\rightarrow [-1,1]^p$ through
$$
n_p(y)\equiv 2^{-p}\sum_{\g=1}^{2^p} e_\g y_\g
\Eq(1.61)
$$
where $e_\g$, $\g=1,\dots , 2^p$ is some enumeration of all $2^p$
vectors in $\R^p$ whose components take values $\pm 1$ only.
 Given $I\subset \N$, we define the set
$D_{|I|}$ as the  set
$$
D_{|I|}\equiv
\left\{ m\in \R^{|I|}\,\mid \exists y\in [-1,+1]^{2^{|I|}}: n_{|I|}(y)=m
\right\}
\Eq(1.62)
$$

\theo{1} {\it Assume that $\limsup_{N\uparrow\infty} \frac
{M} N=0$.   Then for any finite $I\subset \N$ and for all
$0<\b<\infty$, the family of distributions $\QQ_{N,\b}[\o] \circ
\Pi_I^{-1}$ satisfies a  LDP for
almost all $\o\in \O$ with rate function $F_\b$ given by
$$
F_\b^I(\tilde m)=-\sup_{p\in \N}\sup_{{y\in [-1,1]^{2^p}}\atop
{\Pi_In_p(y)=\tilde m}}
\left[\frac 12 \|n_p(y)\|_2^2-\b^{-1}2^{-p}\sum_{\g=1}^{2^p}
I(y_\g)\right]
      +\sup_{y\in \R}\left( \frac 12 y^2-\b^{-1}
I(y)\right)
\Eq(1.7)
$$
where
$$
I(y)\equiv \cases {\frac {1+y}2\ln (1+y)+\frac {1-y}2\ln (1-y)&,
if $|y|\leq 1$\cr
+\infty&, otherwise}
\Eq(1.8)
$$
$F_\b^I$ is lower semi-continuous, Lipshitz-continuous on the interior
of $D_{|I|}$, bounded on $D_{|I|}$ and equal to $+\infty$ on $D_{|I|}^c$.
}


\remark Note that $F_\b^I$ is {\it not } convex in general.

To prove Theorem 1 we will define, for $\tilde m\in \R^I$
$$
F_{N,\b,\e}^I(\tilde m)\equiv -\frac 1{\b N}\ln \mu_{N,\b}[\o]
\left(\|\Pi_I m_N(\s)-\tilde m\|_2\leq \e\right)
\Eq(1.6)
$$
and show that
\item{i)} If $\tilde m\in  D_{|I|}$, then
$$
\lim_{\e\downarrow 0}\lim_{N\uparrow\infty} F_{N,\b,\e}^I(\tilde m)
=F_\b^{I}(\tilde m)
\Eq(1.9)
$$
 almost surely and
\item{ii)} If $\tilde m\in D_{|I|}^c$, then
$$
\lim_{\e\downarrow 0}\lim_{N\uparrow\infty} F_{N,\b,\e}^I(\tilde m)
=+\infty
\Eq(1.91)
$$
almost surely.

{}From these two equations it follows from standard arguments (see e.g. [DZ])
 that for
almost all $\o$
for all Borel-sets $\A\subset \BB(\R^I)$
$$
\eqalign{
-\inf_{\tilde m\in \sminn\A} F_\b^I(\tilde m)
&\leq \liminf_{N\uparrow\infty}\frac 1N\ln \QQ_{N,\b}[\o] \circ
\Pi_I^{-1}\left(\A\right)\cr
&\leq
\limsup_{N\uparrow\infty}\frac 1N\ln \QQ_{N,\b}[\o] \circ
\Pi_I^{-1}\left(\A\right)\leq
-\inf_{\tilde m\in \smcl\A} F_\b^I(\tilde m)
}
\Eq(1.81)
$$
The properties of the rate function will be established directly from its
explicit form \eqv(1.7).


An important feature  is that the rate function is non-random. This means that
under the conditions of the theorem, the thermodynamics of this
disordered system is described in terms of completely deterministic potentials.
{}From the thermodynamic point of view discussed above, this is an
extremely
satisfactory result. Namely in these terms it means that although
the Hamiltonian of our model is a function of an unbounded number
of random macroscopic quantities, we may select any finite subset
of these in which we may be interested and can be assured that
there will exist, with probability one,
in the infinite volume limit, thermodynamic
potentials that are functions of these variable only and which
are, moreover, completely deterministic.
The sole condition for this to hold is that the number of
macroscopic variables goes to infinity with a sublinear rate.

In the remainder of this article we will present the proof of Theorem 1.
There will be three important steps. First, we prove large deviation estimates
for the mass of small balls in $\R^M$, using fairly standard
techniques. The resulting bounds are expressed in terms of a certain random
 function. The crucial step is to show that in a certain strong sense this
function is `self-averaging'. The proof of this fact uses the Yurinskii
martingale representation and exponential estimates (see e.g. [LT]).
These are  finally combined to obtain deterministic estimates on  cylinder
events from which the convergence result \eqv(1.9) then follows rather easily.
\newpage

\chap{2. The basic large deviation estimates}2

In this section we recall exponential upper and lower bounds that have
already been derived in [BGP2]. They provide the starting point of
our analysis.

Let us consider  the quantities

$$
Z_{N,\b,\rho}[\o]( m)\equiv\mu_{N,\b}[\o]\left(\|
m_N(\s)-m\|_2\leq\rho\right)Z_{N,\b}[\o]
\Eq(2.1)
$$
We first proof a large deviation  {\it upper} bound.

\lemma {2.1} {\it
$$
\frac 1{\b N}\ln
Z_{N,\b,\rho}( m)\leq
\Phi_{N,\b}(m)
+\rho(\|t^*\|_2+ \|m\|_2+\rho/2)
\Eq(2.2)
$$
where
$$
\Phi_{N,\b}(m) \equiv\inf_{t\in\R^M} \Psi_{N,\b}(m,t)
\Eq(2.3)
$$
with
$$
\Psi_{N,\b}(m,t)\equiv -(m,t)+\frac 12 \|m\|_2^2+\frac 1{\b N}\sum_{i=1}^N
\ln\cosh\b(\xi_i,t)
\Eq(2.4)
$$
and $t^*\equiv t^*(m)$ is defined through
$
\Psi_{N,\b}(m,t^*(m)) =
\inf_{t\in\R^M} \Psi_{N,\b}(m,t)$,
if such a $t^*$ exists, while otherwise $\|t^*\|\equiv\infty$.}

\proof Note that for arbitrary $t\in\R^M$,
$$
\1_{\left\{\|m_N(\s)-m\|_2\leq \rho\right\}}
\leq\1_{\left\{\|m_N(\s)-m\|_2\leq \rho\right\}}
e^{\b N\left(t,(m_N(\s)-m)\right)+\rho \b N\|t\|_2}
\Eq(2.6)
$$
Thus
$$
\eqalign{
Z_{N,\b,\rho}( m)
& = \E_\s e^{\frac {\b N}2 \|m_N(\s)\|_2^2}
\1_{\left\{\|m_N(\s)-m\|_2\leq \rho\right\}}\cr
&\leq \inf_{{t\in\R^M}}\E_\s
e^{\b N \frac 12 \left(\|m\|_2^2 +2\rho\|m\|_2 +\rho^2\right)}
e^{\b N\left(t, (m_N(\s)-m)\right)+\b N\rho\|t\|_2}\cr
&\leq \inf_{{t\in\R^M}}
e^{\b N \left[\frac 12\|m\|_2^2-(m,t)
+\frac 1{\b N}\sum_{i=1}^N
\ln\cosh\left(\b  (\xi_i,t)\right)\right]}
e^{\b N \rho \left(\|m\|_2 + \|t\|_2+ \rho/2\right)}
}
\Eq(2.7)
$$
This gives immediately the bound of  Lemma 2.1.\endproof

\remark Note that if a finite $t^{*}(m)$ exists, then it is the solution of the
system of equations
$$
m^\mu=\frac 1N\sum_{i=1}^N\xi_i^\mu \tanh\b(\xi_i,t)
\Eq(2.8)
$$

\noindent Next we  prove a corresponding {\it lower} bound.

\lemma{2.2} {\it
For $\rho\geq \sqrt{2\frac MN}$, we have that
$$
\frac 1{\b N}\ln
Z_{N,\b,\rho}( m)\geq \Phi_{N,\b}(m)-\rho(\|m\|_2+\|t^*(m)\|_2-\rho/2)-\frac
{\ln 2}{\b N}
\Eq(2.8bis)
$$
where the notations are the same as in Lemma 2.1.
}

\proof  The technique to prove this bound is the standard one to prove
a Cram\`er-type lower bound (see e.g. [Va]). It is of course enough to
consider the case where $\|t^*\|_2<\infty$.
We define,  for $t^*\in \R^M$, the probability
measures $\tilde \P$ on $\{-1,1\}^N$
through their expectation $\tilde\E_\s$, given by
$$
\tilde \E_\s\bigl(\cdot\bigr)\equiv\frac{\E_\s e^{\b N\left(t^*,m_N(\s)\right)}
\bigl(\cdot\bigr)}{\E_\s e^{\b N\left(t^*,m_N(\s)\right)}}
\Eq(2.9)
$$
We have obviously that
$$
\eqalign{
Z_{N,\b,\rho}(m)&= \tilde\E_\s e^{\frac {\b N}2\|m_N(\s)\|_2^2-
\b N\left(t^*,m_N(\s)\right)}
\1_{\left\{\| m_N(\s)-m\|_2\leq \rho\right\}} \E_\s e^{\b
N\left(t^*,m_N(\s)\right)}
\cr
&\geq e^{-\b N\left(t^*,m\right)-\b N\left(\rho\| t^*\|_2
-\frac 12\|m\|_2^2+\rho\|m\|_2-\rho^2/2\right)}
\E_\s e^{\b N\left(t^*,m_N(\s)\right)}
\tilde \E_\s\1_{\left\{\| m_N(\s)-m\|_2\leq \rho\right\}}\cr
&=e^{\b N\left(\frac 12\|m\|_2^2-(t^*,m)+\frac 1{\b N}
\sum_{i=1}^N\ln\cosh\b(\xi_i,t^*)\right)}
e^{-\b N \rho \left( \|t^*\|_2+\|m\|_2-
\rho/2\right)}\cr
&\times\tilde\P_\s\left[\| m_N(\s)-m\|_2\leq \rho\right]
}
\Eq(2.10)
$$
But, using Chebychev's inequality, we have that
$$
\eqalign{
\tilde\P_\s\left[\|m_N(\s)-m\|_2\leq \rho\right]
&=1-\tilde\P_\s\left[\|m_N(\s)-m\|_2
\geq \rho\right]\cr
&\geq 1-\frac 1{\rho^2} \tilde\E_\s \|m_N(\s)-m\|_2^2
}
\Eq(2.12)
$$
We choose  $t^*(m)$ that satisfies equation \eqv(2.8). Then it is easy to
compute
$$
\tilde\E \|m_N(\s)-
m\|_2^2=\frac MN\left(1-\frac 1N\sum_{i=1}^N \tanh^2(\b(\xi_i,t^*(m)))\right)
\Eq(2.13)
$$
from which the lemma follows. \endproof

In the following lemma we collect a few  properties of $\Phi_{N,\b}(m)$
that arise from convexity. We set
$\G\equiv\left\{m\in\R^M \mid \|t^*(m)\|_2<\infty \right\}$
where $t^*(m)$ is defined in Lemma 2.1,
$D\equiv\left\{m\in\R^M \mid\Phi_{N,\b}(m)>-\infty\right\}$,
and we denote by $\hbox{int}D$ the interior of $D$.
We moreover denote by
$I(x)\equiv \sup_{t\in\R}(tx-\ln\cosh t)$
the Legendre transform of the function $\ln\cosh t$. A simple computation
shows that $I(x)$ coincides with the function defined in \eqv(1.8).

\lemma{2.3} {\it
\item{i)}
$$
\Phi_{N,\b}(m)=\frac{1}{2}\|m\|_2^2-
\inf_{y\in\R^N : m_N(y)=m}
\frac{1}{\b N}\sum_{i=1}^N I(y_i)
\Eq(2.132)
$$
where for each $m\in\R^M$  the infimum is attained or is $+\infty$
vacuously.
\item {ii)}
$$
D=\left\{m\in\R^M \mid \exists y\in [-1, 1]^N s.t.\,\, m_N(y)=m
\right\}
\Eq(2.vero)
$$
\item{iii)} $\Phi_{N,\b}(m)$ is continuous relative to $\hbox{int}D$
\item{iv)} $\G=\hbox{int}D$
\item{v)} If $t^*$ is defined as in Lemma 2.1 and $y^*$ realizes the
infimum in \eqv(2.132), then
$$
\b^2\left(t^*,\frac{\xi^t\xi}N t^*\right)=\frac 1N\sum_{i=1}^N[I'(y^*_i)]^2
\Eq(2.anton)
$$
}

\remark Note that point i) of Lemma 2.3. provides an   alternative
formula for the variational formula \eqv(2.3).

\proof All results of convex analysis used in this proof can be found in
[R]. Note that the function $g(t)\equiv\frac{1}{\b N}\sum_{i=1}^N
\ln\cosh\b(\xi_i,t)$ is a proper convex function on $\R^M$. Denoting by
$h(m)\equiv\sup_{t\in\R^M}\{(m,t)-g(t)\}$ its Legendre transform, it
follows from standard results of convex analysis that $h(m)$ is a
proper convex function on $\R^M$ and that
$$
h(m)=\inf_{y\in\R^N : m_N(y)=m}\frac{1}{\b N}\sum_{i=1}^N I(y_i)
\Eq(2.133)
$$
where for each $m\in\R^M$ the infimum is either attained or is $+\infty$.
This immediately yields i). Denoting by
$\hbox{dom} h\equiv\left\{x\in\R^M \mid h(m)<+\infty\right\}$ the effective
domain
of $h$, we have, by (1.7), that $\hbox{dom} h$ equals the right hand side of
\eqv(2.vero) , and since
$\|m\|_2^2\geq 0$, ii) is proven. iii) simply follows from the
fact that $h$ being convex, it is continuous relative to the interior of
$\hbox{dom} h$. Finally, to prove iv), we will make use of the following two
important results of convex analysis. First, the subgradient of $h$ at
$m$, $\partial h(m)$, is a non empty set if and only if $m$ belongs to
the interior of $\hbox{dom} h$,
i.e., $m\in \hbox{int}D$. $\partial h(m)$ is moreover a
bounded convex set. Next, $(m,t)-g(t)$ achieves its supremum at
$t^*\equiv t^*(m)$ if and only if $t^*\in \partial h(m)$.
To prove v) we only have to consider the case where $t^*$ exists and
consequently $|y^*_i|<1$ for all $i$.
Using the fact that $I'(x)=\tanh^{-1}(x)$ and the definition of $I(x)$ as the
Legendre transform of $\ln\cosh(t)$, formula \eqv(2.anton) follows from a
simple computation.
This concludes
the proof of the lemma.
\endproof

We see that as long as $\rho$ can be chosen as a function of $N$
that tends to zero as $N$  goes to infinity, Lemma 2.1 and Lemma 2.2
seem to provide asymptotically coinciding upper and lower bounds,
at least for such
$m$ for which $t^*(m)$ is bounded. The unpleasant feature in these bounds is
that $\Psi_{N,\b}$ is a rather complicated random function and that the
$\Phi_{N,\b}$ is defined through an infimum of such a function. In the next
section we analyse this problem and show that this function
 is essentially non-random.
\newpage

\chap{3. Self averaging }3

We show now that the random upper and lower bounds derived in the
last section are actually  with large probability independent of the
realization of the randomness.
 In fact we will prove
this under the restriction that $m$ should be such that, at least
on a subspace of full measure, $t^*(m)$ has a uniformly bounded
$\ell_2$-norm. With this in mind the result will follow from the
next proposition.
Let in the sequel $\O_1\subset \O$ denote the subspace for which
$\|\xi^t[\o]\xi[\o]/N\|=
\|\xi[\o]\xi^t[\o]/N\|\leq (1+\sqrt\a)^2(1+\e)$ holds
for some fixed small $\e$ ($\e=1$ will be a suitable choice).
We recall from [ST,BG1,BGP1] that
$\P[\O_1]\geq 1-4Ne^{-\e N^{1/6}}$.

\proposition {3.1} {\it For any $R<\infty$ there exists $0<\d<1/2$ and
 a set $\O_2\subset \O$
with  $\P[\O_2]\geq 1-
e^{-N\a^{1-2\d}/R}$, such that for all $\o\in \O_1\cap\O_2$,
$$
\sup_{t:\,\|t\|_2\leq R}
\left|\Psi[\o](m,t)-\E\Psi(m,t)\right|\leq
\a^{1/2-\d}(6+2\|m\|_2)
\Eq(3.1)
$$
}

\remark The subspace $\O_2$ does {\it not} depend on $m$.

 Note that an immediate corollary to Proposition 3.1 is that, under its
assumptions,
$$
\left|\inf_{t:\,\|t\|_2\leq R }\Psi[\o](m,t)-\inf_{t:\,\|t\|_2\leq R}
\E\Psi(m,t)\right|\leq
\a^{1/2-\d}(6+2\|m\|_2)
\Eq(3.02)
$$

\remark An obvious consequence of \eqv(3.02) is the
observation that if
$m\in \R^M$ and $\o\in \O_1\cap\O_2$ are such that
$$
\inf_{t\in \R^M} \Psi[\o](m,t)=\inf_{t\,:\|t\|_2\leq R} \Psi[\o](m,t)
\Eq(3.2)
$$
and
$$
\inf_{t\in \R^M} \E\Psi(m,t)=\inf_{t\,:\|t\|_2\leq R} \E\Psi[\o](m,t)
\Eq(3.2a)
$$
then
$$
\left|\Phi[\o](m)-\inf_{t}
\E\Psi(m,t)\right|\leq c\a^{1/2-\d}
\Eq(3.1bis)
$$

\proof
The proof of the proposition follows from the fact that
for bounded values of $t$, $\Psi(m,t)$ differs uniformly only
little from its expectation. This will be proven by first
controlling a lattice supremum, and then using some a priori
Lipshitz bound on $\Psi(m,t)$. We prove the Lipshitz bound first.

\lemma {3.2} {\it Assume that $\o\in \O_1$.
Then
$$
\left|\Psi[\o](m,t)-\Psi[\o](m,s)\right|\leq
\left((1+\sqrt\a)(1+\e) +\|m\|_2\right)\|t-s\|_2
\Eq(3.3)
$$
}

\proof Note that
$$
\eqalign{
\left|\Psi(m,t)-\Psi(m,s)\right|&\leq
\left| -(m, t-s)+\frac 1{\b N} \sum_i\left[\ln\cosh(\b(\xi_i,t))-
\ln\cosh(\b(\xi_i,s))\right]\right|\cr
&\leq \|m\|_2\|t-s\|_2 +\left| \frac 1{\b N}
 \sum_i\left[\ln\cosh(\b(\xi_i,t))-
\ln\cosh(\b(\xi_i,s))\right]\right|
}
\Eq(3.4)
$$
On the other hand, by the mean-value theorem,
there exists $\tilde t$ such that
$$
\eqalign{
&\left| \frac 1{\b N}
 \sum_i\left[\ln\cosh(\b(\xi_i,t))-
\ln\cosh(\b(\xi_i,s))\right]\right|=
\left|\left(t-s,\frac 1N\sum_i\xi_i\tanh(\b(\xi_i,\tilde
t))\right)\right|\cr
&=\left|\frac 1N \sum_i(t-s,\xi_i)\tanh(\b(\xi_i,\tilde t))\right|
}
\Eq(3.5)
$$
Using the Schwartz inequality, we have that
$$
\eqalign{
&\left|\frac 1N \sum_i(t-s,\xi_i)\tanh(\b(\xi_i,\tilde t))\right|
\leq \frac 1N
\sqrt{\sum_i(t-s,\xi_i)^2}\sqrt{\sum_i\tanh^2(\b(\xi_i,\tilde
t))}\cr
&\leq \sqrt{\left((s-t,\sum_i\frac {\xi^t_i\xi_i}N
(s-t))\right)}\cr
&\leq \sqrt{\left\|\frac {\xi^t\xi}N\right\|}\|t-s\|_2
}
\Eq(3.6)
$$
But this implies the lemma.\endproof

Let us now introduce a lattice $\WW_{N,M}$ with spacing $1/\sqrt
N$ in $\R^M$. We also denote by $\WW_{N,M}(R)$ the intersection
of this lattice with the ball of radius $R$. The point is that
first, for any $t\in \R^M$, there exists a lattice point $s\in
\WW_{N,M}$ such that $\|s-t\|_2\leq \sqrt \a$, while on the other
hand
$$
\left|\WW_{N,M}(R)\right| \leq e^{\a N(\ln (R/\a))}
\Eq(3.7)
$$

\lemma {3.3} {\it
$$
\P\left[\sup_{t\in \WW_{N,M}(R)}|\Psi(m,t)-\E\Psi(m,t)|>x\right]
\leq e^{-N\left(\frac {x^2}{R}(1-\frac {1}{2}e^{x/R})-\a\ln(R/\a)\right)}
\Eq(3.8)
$$
}

\proof Clearly we only have to prove that for all $t\in
\WW_{N,M}(R)$
$$
\P\left[|\Psi(m,t)-\E\Psi(m,t)|>x\right]
\leq e^{-N\frac {x^2}{R}(1-\frac {1}{2}e^{x/R})}
\Eq(3.9)
$$
To do this we write $\Psi(m,t)-\E\Psi(m,t)$
as a sum of martingale differences and use an exponential Markov
inequality for martingales. Note first that
$$
\Psi(m,t)-\E\Psi(m,t)=\frac 1{\b N}\sum_{i=1}^N
\left(\ln\cosh(\b(\xi_i,t))-\E\ln\cosh(\b(\xi_i,t))\right)
\Eq(3.10)
$$

We  introduce the decreasing sequence of sigma-algebras
$\FF_{k,\k}$ that are generated by the random variables
$\left\{\xi_i^\mu\right\}^{1\leq\mu\leq M}_{i\geq k+1}\cup
\left\{\xi_k^\mu\right\}^{\mu\geq \k}$.
We set
$$
\tilde f^{(k,\k)}_N\equiv \E\left[
\b^{-1}\sum_i\ln\cosh(\b(\xi_i,t))         \big|\FF_{k,\k}\right]-\E\left[
\b^{-1}\sum_i\ln\cosh(\b(\xi_i,t))\big|\FF^+_{k,\k}\right]
\Eq(3.11)
$$
where for notational convenience we have set
$$
\FF^+_{k,\k}=\cases{\FF_{k,\k+1},&if $\k<M$\cr
\FF_{k+1,1}&if $\k=M$}
\Eq(3.12)
$$
Notice that we have the identity
$$
\Psi(m,t)-\E\Psi(m,t)
\equiv \frac 1N\sum_{k=1}^N
\sum_{\k=1}^M\tilde f^{(k,\k)}_N
\Eq(3.13)
$$
Our aim is to use an exponential Markov inequality for
martingales. This requires in particular bounds on the
conditional Laplace transforms of the martingale differences (see
e.g. [LT]).
Namely, we clearly have that
$$
\eqalign{
&\P\left[ \left|\sum_{k=1}^N\sum_{\k=1}^M
\tilde f^{(k,\k)}_N\right|\geq N x\right] \leq
 2\inf_{u\in\R} e^{-|u|Nx}\E\exp\left\{u\sum_{k=1}^N
\sum_{\k=1}^M
\tilde f^{(k,\k)}_N\right\}\cr
&\quad=2\inf_{u\in R} e^{-|u|Nx}
\E\left[\E\left[\dots\E\left[e^{u\tilde f^{(1,1)}_N}\big |
\FF_{1,1}^+\right] e^{u\tilde f^{(1,2)}_N}\big |
\FF_{1,2}^+\right]\dots e^{u\tilde f^{(N,M)}_N}\big |
\FF^+_{N,M}\right]
}
\Eq(3.14)
$$
Now notice that
$$
\eqalign{
\tilde f_N^{(k,\k)}&=
\E[\b^{-1}\sum_i \ln\cosh(\b(\xi_i,t))|\FF_{k,\k}]-
\E[\b^{-1}\sum_i \ln\cosh(\b(\xi_i,t))|\FF^+_{k,\k}]\cr
&=\E[\b^{-1} \ln\cosh(\b(\xi_k,t))|\FF_{k,\k}]-
\E[\b^{-1} \ln\cosh(\b(\xi_k,t))|\FF^+_{k,\k}]\cr
&=\E[\b^{-1}
\ln\cosh(\b\left(\sum_{\mu\neq\k}\xi^\mu_kt_\mu+\xi^\k_kt_\k\right))|\FF_{k,\k}]-
\E[\b^{-1}
\ln\cosh(\b\left(\sum_{\mu\neq\k}\xi^\mu_kt_\mu+\xi^\k_kt_\k\right))|\FF^+_{k,\k}]\cr
&=\frac 12 \b^{-1}\E\left[
\ln\cosh(\b\left(\sum_{\mu\neq\k}\xi^\mu_kt_\mu+\xi^\k_kt_\k\right))
-\ln\cosh(\b\left(\sum_{\mu\neq\k}\xi^\mu_kt_\mu-\xi^\k_kt_\k\right))
\big|\FF_{k,\k}\right]
}
\Eq(3.15)
$$
Now we use the fact that
$$
\frac {\cosh(a+b)}{\cosh(a-b)}=\frac {1+\tanh a\tanh b}{1-\tanh
a\tanh b}
\leq \frac {1+ \tanh|b|}{1-\tanh|b|}\leq e^{2|b|}
\Eq(3.16)
$$
to deduce from \eqv(3.15) that
$$
|\tilde f_N^{(k,\k)}|\leq |t_\k|
\Eq(3.17)
$$
Using the standard  inequalities
$
e^x\leq 1+x+\frac {x^2}2 e^{|x|}$ and
$1+y\leq e^y$
we get therefore
$$
\E\left[e^{u
\tilde f^{(k,\k)}_N(\tilde m)}\big|\FF_{k,\k}^+\right]\leq
\exp\left(\frac {u^2}2 t_\k^2
e^{|u||t_\k|}\right)
\Eq(3.18)
$$
{}From this and \eqv(3.14) we get now
$$
\eqalign{
\P\left[|\Psi(m,t)-\E\Psi(m,t)|>x\right]&\leq 2\inf_{u}e^{-uNx+
\frac {u^2}2 N\|t\|_2^2
e^{|u|\|t\|_\infty}}\cr
&\leq\cases{ 2e^{-N \frac {x^2}{\|t\|_2^2}
\left(1-\frac {1}{2}
e^{x/\|t\|_2}\right)},&if  $\|t\|_2\geq 1$\cr
 2 e^{-N x^2\left(1-\frac {1}{2}e^{x}\right)},&if $\|t\|_2< 1$\cr
}
}
\Eq(3.19)
$$
where the last inequality is obtained by choosing $u=x/\|t\|_2^2$
in the first and $u=x/\|t\|_2$ in the second case. This gives the
lemma. \endproof

We can now continue the proof of Proposition 3.1. Choose
$0<\d<1/2$ and define $\O_2$ to be  the set of $\o\in \O$ for
which
$$
\sup_{t\in \WW_{N,M}(R)}|\Psi(m,t)-\E\Psi(m,t)|\leq \a^{1/2-\d}
\Eq(3.20)
$$
By Lemma 3.3,
$$
\eqalign{
\P[\O_2]&\geq 1-\exp\left(-N\frac {\a^{1-2\d}}{R}
(1-\frac {1}{2}e^{\a^{1/2-\d}/R})+N\a\ln (R/\a)\right)\cr
&=1-\exp\left(-NO(\a^{1-2\d}/R)\right)
}
\Eq(3.21)
$$
Combining Lemma 3.2 with \eqv(3.20) and taking into account the
remark preceeding Lemma 3.3, we see that on $\O_1\cap\O_2$,
$$
\sup_{t\,:\|t\|_2\leq R}|\Psi(m,t)-\E\Psi(m,t)|\leq
\a^{1/2-\d}+2\sqrt{\a}(\|m\|_2+(1+\sqrt\a)(1+\e))
\leq \a^{1/2-\d}(6+\|m\|_2)
\Eq(3.22)
$$
for $\a$ small, which proves Proposition 3.1.\endproof
\newpage

\chap {4. Proof of the Theorem}4

The results of Sections 2.1 and 3.1 can now be combined to get a
large deviation principle in the product topology. The point here is that,
apart from the possibility that $t^*(m)$ may become unbounded, the
estimates in Lemma 2.1 and Lemma 2.2 together with Proposition 3.1 tell
us that up to corrections that tend to zero with $N$, the quantity
$(\b N)^{-1}\ln Z_{N,\b,\sqrt {2\a}}(m)$
is given by the
infimum over $t$ of the completely non-random function
$\E\Psi_{N,\b}(m,t)$. We will first prove that for all
$\tilde m\in D_I$ \eqv(1.9) holds. The main step in the proof of this fact is
 the following theorem.

\theo {4.1} {\it       Assume that $\limsup_{N\uparrow\infty} \frac
{M(N)} N=0$ and that $0<\b<\infty$.
Then there exists a set $\tilde \O\subset \O$ with $\P[\tilde\O]=1$
such that for all finite subsets $I\subset\N$ and for all
$\tilde m\in [-1,1]^I$ such that for all $\e>0$ there exists
$c=c(\tilde m, \e)<\infty$, $\exists N_0\leq \infty$,$\forall N\geq N_0$,
$$
\sup_{m:\,\|\Pi_I m-\tilde m\|_2\leq\e}\inf_{t\in \R^M} \E \Psi_{N,\b}(m,t)=
\sup_{m:\,\|\Pi_I m-\tilde m\|_2\leq \e}\inf_{t\in \R^M:\,\|t\|_2\leq c}
\E\Psi_{N,\b}(m,t)
\Eq(4.1)
$$
it holds that for all $\o\in\tilde \O$,
$$
\eqalign{
&\lim_{\e\downarrow 0}\lim_{N\uparrow\infty}
F_{N,\b,\e}^{I}[\o](\tilde m)\cr
&=-\sup_{p\in \N}\sup_{{y\in [-1,1]^{2^p}}\atop{\Pi_I n_p(y)=\tilde m}}
\left[\frac 12 \|n_p(y)\|_2^2-\b^{-1}2^{-p}\sum_{\g=1}^{2^p}
I(y_\g)\right] + \sup_{y\in\R}\left(\frac{1}{2}y^2-\b^{-1}I(y)\right)
}
\Eq(4.2)
$$
}

\remark The assumption in Theorem 4.1 looks horrible at first glance. The
reader will observe that it is made in order to allow us to apply the
self-averaging results from the last section. We will show later, however,
that the set of values $\tilde m$ for which it is satisfied
can be constructed explicitly and is nothing else than $D_{|I|}$.

\proof We will first establish an upper bound for the quantity
$$
Z_{N,\b,\e}^{I}[\o](\tilde m) \equiv
\mu_{N,\b}[\o]\left(\|\Pi_I m_N(\s)-\tilde m\|_2\leq\e\right)
Z_{N,\b}[\o]
\Eq(4.0)
$$
To do so, notice that on $\O_1$,
$\|m_N(\s)\|_2\leq (1+\sqrt\a)\sqrt {(1+\e)}<2$ for all $\s$. We may cover
the ball of radius 2 with balls of radius $\rho\sim\sqrt{\a}$, centered
at the lattice points in $\WW_{N,M}(2)$.
We then have that on $\O_1$,
$$
\eqalign{
 Z_{N,\b,\e}^{I}[\o](\tilde m)&\leq \sum_{{m\in \WW_{N,M}(2)}\atop
{\|\Pi_I m-\tilde m\|_2\leq \e}}
Z_{N,\b,\rho}[\o](m)\cr
&\leq \sup_{{m\in \WW_{N,M}(2)}\atop
{\|\Pi_I m-\tilde m\|_2\leq \e}}Z_{N,\b,\rho}[\o](m)\sum_{{m\in
\WW_{N,M}(2)}\atop
{\|\Pi_I m-\tilde m\|_2\leq \e}}1\cr
&\leq \sup_{{m:\, \|m\|_2<2}
\atop{\|\Pi_I m-\tilde m\|_2\leq \e}
}Z_{N,\b,\rho}[\o](m) e^{\a N(\ln 2/\a)}
}
\Eq(4.3)
$$
As long as $\a\downarrow 0$, the factor $e^{\a N(\ln2/\a)}$ in the
upper bound is irrelevant for the exponential asymptotic, as is the
difference between $\e$ and $\e-\rho$.
Using the estimates used in the proof of  Lemma 2.1, we can
replace $Z_{N,\b,\rho}[\o](m)$ in \eqv(4.3)
by its upper  bound in terms of the function $\Psi$. Namely,
$$
\eqalign{
\frac 1{\b N}\ln Z_{N,\b,\e}^I[\o](\tilde m)&\leq
\sup_{{m:\, \|m\|_2<2}\atop{\|\Pi_I m-\tilde m\|_2\leq \e}}
\inf_{{t\in\R^M}\atop
{\|t\|_2\leq c}} \Psi_{N}[\o](m,t) +\rho (c+2+\rho/2)
+ \b^{-1}\a\ln 1/\a\cr
}
\Eq(4.5)
$$
 Finally, combining \eqv(4.5) with \eqv(3.02), we get that, for
$\o\in\O_1\cap\O_2$ and for any $c$,
$$
\frac 1{\b N}\ln Z_{N,\b,\e}^I[\o](\tilde m)\leq
\sup_{m:\,\|\Pi_I m-\tilde m\|_2\leq \e}\inf_{{t\in\R^M}\atop
{\|t\|_2\leq c}}\E \Psi_{N}(m,t)+10\a^{1/2-\d} +\rho (c+2+\rho/2)
+ \b^{-1}\a\ln 1/\a
\Eq(4.9)
$$
By assumption, there exists a value $c<\infty$, such that the
true minimax over $\E\Psi_N(m,t)$ is taken for a value of $t$
with norm bounded uniformly in $N$ by some constant $c$. The
constant $c$ in \eqv(4.9) is then chosen as this same constant,
and then the restriction $\|t\|_2\leq c$ is actually void, and
the minimax is taken for some values $(m^*, t^*)$ which depend
only on $\tilde m$ and $\e$.
This is already essentially the desired form of the
upper bound.

We now turn to the more subtle problem of obtaining the corresponding
form of the lower bound.
 Trivially,
$$
Z_{N,\b,\e+\rho}^{I}[\o](\tilde m)\geq Z_{N,\b,\rho}[\o](m^*)
\Eq(4.10)
$$
We will modify slightly the derivation of the lower bound
for $Z_{N,\b,\rho}[\o](m^*)$. Namely, instead of
defining the shifted measure $\tilde P$ with respect to the
random value of $t$ that realizes the infimum of
$\Psi_N[\o](m^*,t)$, we do this with the deterministic value
$t^*$ that realizes the infimum of $\E\Psi_N(m^*,t)$.
This changes nothing in the computations in \eqv(2.10) and
\eqv(2.12). What changes, is however the estimate on
$\tilde\E_\s\|m_N(\s)-m^*\|_2^2$, since $t^*$ does not
satisfy \eqv(2.8) but is instead solution of the equations
$$
m^*_\mu=\E \xi^\mu_1\tanh(\b(\xi_1,t^*))
\Eq(4.12)
$$
Thus in place of \eqv(2.13) we get
$$
\eqalign{
&\tilde\E_\s \|m_N(\s)-m^*\|_2^2=\cr
&\frac{\E_\s \prod_{i=1}^Ne^{\b (t^*,
\xi_i\s_i)}\sum_{\nu}\left(
N^{-2}\sum_{j,k}\xi_j^\nu\xi_k^\nu\s_j\s_k -2m^*_\nu N^{-1}\sum_j
\xi_j^\nu\s_j+(m^*_\nu)^2\right)}{\prod_{i=1}^N\cosh\b (\xi_i,t^*)}\cr
&=\frac 1{N^2}\sum_\nu\sum_j 1+\frac 1{N^2}
\sum_{\nu}\sum_{j\neq k}\tanh(\b (t^*,\xi_j))
\tanh(\b (t^*,\xi_k))\xi_j^\nu\xi_k^\nu\cr
&
-2\frac 1N\sum_{j}\sum_{\nu} m^*_\nu\tanh(\b (t^*,\xi_j))\xi_j^\nu
+\sum_\nu (m^*_\nu)^2\cr
&=\frac MN\left(
1-\frac 1N\sum_i\tanh^2(\b(t^*,\xi_i))\right)+
\sum_\nu\left(\frac 1N \sum_i \xi_i^\nu\tanh(\b(t^*,\xi_i))-m^*_\nu\right)^2
}
\Eq(4.11)
$$
The first summand in \eqv(4.10) is bounded by $\a$, and we have
to control the second. To do so we use \eqv(4.12) to write
$$
\eqalign{
&\sum_\nu\left(\frac 1N \sum_i
\xi_i^\nu\tanh(\b(t^*,\xi_i))-m^*_\nu\right)^2\cr
=&\sum_\nu\left(\frac 1N \sum_i
\xi_i^\nu\tanh(\b(t^*,\xi_i))- \E \xi^\nu_1\tanh(\b(\xi_1,t^*))
\right)^2\cr
=&\sum_\nu\left(\frac 1N \sum_i
\xi_i^\nu\tanh(\b(t^*,\xi_i))- \E\frac 1N \sum_i
\xi_i^\nu\tanh(\b(t^*,\xi_i))\right)^2
\cr
&\equiv G_N(t^*)
}
\Eq(4.13)
$$
We will now prove, in analogy to Proposition 3.1, that $G_N(t)$
is  actually small with large probability. This will be slightly more
 complicated than in Proposition 3.1 and will, in fact consist of two
steps. The first is a fairly crude bound on $G_N(t)$ that in a second step
will be used to obtain a refined one.

\lemma {4.2} {\it  For all $\o\in \O_1$,
$$
G_N[\o](t)\leq 6
\Eq(4.14)
$$
}

\proof Let us for notational simplicity set $T_i\equiv \tanh(\b(\xi_i,t))$.
We have that
$$
\eqalign{
G_N(t)&\leq 2\sum_{\mu=1}^M\left(\left[\frac 1N\sum_i\xi_i^\mu T_i\right]^2
+\left[\frac 1N\sum_i\E\xi_i^\mu T_i\right]^2\right)\cr
&=\frac 2{N^2}\sum_{\mu=1}^M \sum_{i,j} \left(\xi_i^\mu\xi_j^\mu T_iT_j
+\E(\xi_i^\mu T_i)\E(\xi_j^\mu T_j)\right)\cr
}
\Eq(4.141)
$$
For the first term, we can use simply that
$$
\frac 2{N^2}\sum_{\mu=1}^M \sum_{i,j} \xi_i^\mu\xi_j^\mu T_iT_j
\leq 2\left\|\frac {\xi \xi^t}N\right\| \left(\frac 1N
\sum_iT_i^2\right)
\leq  2\left\|\frac {\xi \xi^t}N\right\|
\Eq(4.142)
$$
But on $\O_1$, the norm in the last line is bounded by $(1+\sqrt\a)^2(1+\e)$.
To bound the second term in \eqv(4.141), we use the independence of
both $\xi^\mu_i$ and $T_i$ for different indices $i$ to write
$$
\eqalign{
&\frac 2{N^2}\sum_{\mu=1}^M \sum_{i,j}\E(\xi_i^\mu T_i)\E(\xi_j^\mu T_j)=
\frac 2{N^2}\sum_{\mu=1}^M \sum_{i,j}
\E\left(\xi_i^\mu T_i\xi_j^\mu T_j\right)\cr
&+\frac 2{N^2}\sum_{\mu=1}^M \sum_i
\left( (\E\xi_i^\mu T_i)^2-\E (T_i)^2\right)\cr
&\leq 2\E \left\|\frac {\xi \xi^t}N\right\| +\frac {2M}N
\cr
&\leq 2\a + 2(1+\sqrt\a)^2(1+\e')
}
\Eq(4.143)
$$
Combining these two bounds we get \eqv(4.14).\endproof

Lemma 4.2 tells us that $G_N(t)$ is bounded, but not yet that it is small.
To do this, we observe first that its mean value is small.

\lemma {4.3} {\it
$$
0\leq \E G_N(t)\leq \a
\Eq(4.211)
$$
}

\proof
$$
\eqalign{
\E G_N(t)&=\sum_{\nu=1}^M\E \left[\frac 1N\sum_i\xi_i^\nu
\tanh(\b(t,\xi_i))- \E\frac 1N \sum_i
\xi_i^\nu\tanh(\b(t,\xi_i))\right]^2\cr
&=\sum_{\nu=1}^M \left(\E\left[\frac 1N\sum_i\xi_i^\nu
\tanh(\b(t,\xi_i))\right]^2-\left[\frac 1N\sum_i\E \xi_i^\nu
\tanh(\b(t,\xi_i))\right]^2\right)\cr
&=\sum_{\nu=1}^M \left(\frac 1{N^2}\sum_i\E
\tanh^2(\b(t,\xi_i))-\frac 1{N^2}\sum_i\left[\E \xi_i^\nu
\tanh(\b(t,\xi_i))\right]^2\right)\cr
&\leq \frac MN
}
\Eq(4.212)
$$
where we have used the independence of the summands for different
indices $i$.
\endproof

In the sequel we will need that the mean value of $G_N(t)$ does not differ
much from its conditional expectation relative to $\O_1$. Namely,
$$
|\E G_N(t) - \E [G_{N}(t)|\O_1]|\leq 2M e^{-\e N^{1/6}}
\Eq(4.213)
$$
is arbitrarily small.

Finally, we will show that on $\O_1$, with large probability,
$G_N(t)$ differs only little from
its conditional expectation relative to $\O_1$.

\lemma{4.4} {\it  Assume that $x\gg (\ln N)/\sqrt N$. Then,
$$
\P\left[|G_N(t)-\E [G_N(t)|\O_1]|\geq x\big|\O_1 \right]
\leq e^{-b\sqrt N x}
\Eq(4.15)
$$
for some positive constant $b$.
}

\proof Basically the proof of this lemma relies on the same technique as
that of Proposition 3.1. However, a number of details are modified. In
particular, we use a coarser filtration of $\FF$ to define our martingale
differences. Namely, we denote by
$
\FF_k
$
the sigma algebra generated by the random variables
$\{\xi_i^\mu\}_{i\geq k}^{\mu\in \N}$.
 We also introduce the
trace sigma algebra $\tilde\FF\equiv \FF\cap\O_1$
and by $\tilde\FF_k\equiv\FF_k\cap\O_1$
the corresponding filtration of the trace sigma algebra. We set
$$
f_N^{(k)}\equiv\E\left[G_N(t)|\tilde\FF_k\right]-
\E\left[G_N(t)|\tilde\FF_{k+1}\right]
\Eq(4.16)
$$
Obviously, we have for $\o\in \O_1$
$$G_N[\o](t)-\E [G_N(t)|\O_1]=\sum_{k=1}^N f_N^{(k)}
\Eq(4.17)
$$
Thus the lemma will be proven if we can prove an estimate of the
form \eqv(4.15) for the sum of the $f_N^{(k)}$.
This goes just as in the proof of Proposition 3.1, i.e. relies on uniform
bounds on the conditional Laplace transforms
$$
\E\left[e^{u f_N^{(k)}}\big|\tilde \FF_{k+1}\right]
\Eq(4.18)
$$
The strategy to get those is very similar to the one used in [BGP3] and [B].
We introduce
$$
G_N^{(k)}(t,z)\equiv \sum_\mu\left(\frac 1N \sum_{i\neq k}
\xi_i^\mu T_i- \E\frac 1N \sum_i
\xi_i^\mu T_i +\frac zN \xi_k^\mu T_k\right)^2
\Eq(4.19)
$$
and set
$$
g_k(z)\equiv G_N^{(k)}(t,z)-G_N^{(k)}(t,0)
\Eq(4.20)
$$
We then have that
$$
f_N^{(k)}=\E\left[g_k(1)|\tilde \FF_k\right]-\E\left[g_k(1)|\tilde \FF_{k+1}
\right]
\Eq(4.21)
$$
since $G_N^{(k)}(t,0)$ is independent of the random variables $\xi_k$.
On the other hand,
$$
g_k(1)=\int_0^1dz\,g'_k(z)
\Eq(4.22)
$$
and
$$
g'_k(z)=2
\sum_{\nu=1}^M \left[\frac 1N\sum_{i\neq k}\xi_i^\nu
T_i- \E\frac 1N \sum_i
\xi_i^\nu T_i+\frac zN \xi_k^\nu T_k\right] \frac 1N\xi_k^\nu T_k
\Eq(4.220)
$$
Let us first get a uniform bound on $|f_N^{(k)}|$ on $\O_1$.
{}From the formulas above it follows clearly that
$$
|f_N^{(k)}| \leq 2\sup_z |g'_k(z)|
\Eq(4.221)
$$
But using the Schwartz inequality,
$$
\eqalign{
|g'_k(z)|&\leq \frac 2N \sum_\mu \left|\frac 1N\sum_{i\neq k}\xi_i^\nu
T_i- \E\frac 1N \sum_i
\xi_i^\nu T_i+\frac zN \xi_k^\mu T_k\right|\cr
&\leq \frac 2N \sqrt M \sqrt{\sum_\mu \left[
\frac 1N\sum_{i\neq k}\xi_i^\nu
T_i- \E\frac 1N \sum_i
\xi_i^\nu T_i+\frac zN \xi_k^\mu T_k\right]^2}\cr
&=\frac {2\sqrt M}N \sqrt{G_N^{(k)}(t,z)}
}
\Eq(4.222)
$$
But on $\O_1$ it is trivial to check that $G_N^{(k)}(t,z)$ satisfies,
for $z\in [0,1]$, the same bound as $G_N(t)$. So that on $\O_1$,
$$
|g'_k(z)|\leq\frac {12\sqrt M}N
\Eq(4.223)
$$
Now we turn to the estimation of the conditional Laplace transform.
Using the standard inequality
$$
e^x\leq 1+x+\frac 12 x^2e^{|x|}
\Eq(4.224)
$$
we get
$$
\eqalign{
\E\left[e^{u f_N^{(k)}}\big|\tilde \FF_{k+1}\right]
&\leq 1 + \frac 12 u^2 \E\left[\left(f_N^{(k)}\right)^2 e^{|u||f_N^{(k)}|}
\big|\tilde \FF_{k+1}\right]\cr
&1+  \frac 12 u^2 e^{|u|\frac {12\sqrt M}N} \E\left[\left(f_N^{(k)}\right)^2
\big|\tilde \FF_{k+1}\right]
}
\Eq(4.225)
$$
A simple computation (see [BGP3]) shows that
$$
\eqalign{
 \E\left[\left(f_N^{(k)}\right)^2
\big|\tilde \FF_{k+1}\right]&\leq
 4\E\left[\left( g_k(1)\right)^2
\big|\tilde \FF_{k+1}\right]\cr
&=4\E\left[\left( \int_0^1 dz\, g'_k(z)\right)^2
\big|\tilde \FF_{k+1}\right]\cr
&\leq 4\E\left[ \int_0^1 dz\,\left(g'_k(z)\right)^2
\big|\tilde \FF_{k+1}\right]\cr
&\leq 4\sup_{0\leq z\leq 1}\E\left[ \left(g'_k(z)\right)^2
\big|\tilde \FF_{k+1}\right]\cr
}
\Eq(4.226)
$$
Let us write
$$
\eqalign{
g'_k(z)&=2
\sum_{\nu=1}^M \left[\frac 1N\sum_{i}\xi_i^\nu
T_i- \E\frac 1N \sum_i
\xi_i^\nu T_i\right] \frac 1N\xi_k^\nu T_k\cr
&+2\sum_{\nu=1}^M \frac {z-1}{N^2} T_k^2
}
\Eq(4.227)
$$
Thus
$$
\eqalign{
\left(g'_k(z)\right)^2&\leq 8 \left(
\sum_{\nu=1}^M \left[\frac 1N\sum_{i}\xi_i^\nu
T_i- \E\frac 1N \sum_i
\xi_i^\nu T_i\right] \frac 1N\xi_k^\nu T_k\right)^2\cr
&+8 T_k^4(z-1)^2\frac {M^2}{N^4}
}
\Eq(4.228)
$$
Let us abbreviate the two summands in \eqv(4.228) by (I) and (II).
The term (II) is of order $\a^2 N^{-2}$ and thus can simply be bounded
uniformly. We have to work a little more to control the conditional
expectation of the  first. We write
$$
\eqalign{
&\E\left[ (I)
\big|\tilde \FF_{k+1}\right]\cr
&=\frac 8{N^2}\E\left[\sum_{\mu,\nu} \xi_k^\mu\xi_k^\nu T_k^2
\left[\frac 1N \sum_{i}
\xi_i^\mu T_i- \E\frac 1N \sum_i
\xi_i^\mu T_i \right]
\left[
\frac 1N \sum_{i}
\xi_i^\nu T_i- \E\frac 1N \sum_i
\xi_i^\nu T_i \right]
\big|\tilde \FF_{k+1}\right]\cr
}
\Eq(4.229)
$$
We observe that under the expectation conditioned on $\tilde\FF_{k+1}$ we
may interchange the indices of $1\leq j\leq k$ and use this to
symmetrize the expression \eqv(4.229).(Notice that this is the reason
why we separated the $z$-dependent contribution in \eqv(4.228)).This gives
$$
\eqalign{
&\E\left[ (I)
\big|\tilde \FF_{k+1}\right]\cr
&=\frac 8{N^2}\E\left[\sum_{\mu,\nu} \sum_{j=1}^k\frac {\xi_j^\mu\xi_j^\nu}k
 T_j^2
\left[\frac 1N \sum_{i}
\xi_i^\mu T_i- \E\frac 1N \sum_i
\xi_i^\mu T_i \right]
\left[
\frac 1N \sum_{i}
\xi_i^\nu T_i- \E\frac 1N \sum_i
\xi_i^\nu T_i \right]
\big|\tilde \FF_{k+1}\right]\cr
&\leq \frac 8{N^2}\E\left[\left\| \sum_{j=1}^k\frac {\xi_j\xi_j^t}k
 T_j^2\right\| \sum_{\mu=1}^M \left[\frac 1N \sum_{i}
\xi_i^\mu T_i- \E\frac 1N \sum_i
\xi_i^\mu T_i \right]^2\Big|\tilde \FF_{k+1}\right]
}
\Eq(4.230)
$$
But by Lemma 4.2, on $\O_1$,
$$
\sum_{\mu=1}^M \left[\frac 1N \sum_{i}
\xi_i^\mu T_i- \E\frac 1N \sum_i
\xi_i^\mu T_i \right]^2=G_N(t)\leq 6
\Eq(4.231)
$$
and since
$$
\left\| \sum_{j=1}^k\frac {\xi_j\xi_j^t}k
 T_j^2\right\| \leq
\left\| \sum_{j=1}^k\frac {\xi_j\xi_j^t}k\right\|\equiv \|B(k)\|
\Eq(4.232)
$$
we get that
$$
\E\left[ (I)
\big|\tilde \FF_{k+1}\right]\leq \frac{48}{N^2}
\E\left[\|B(k)\|\,\big|\O_1\right]\leq \frac{48}{N^2}\E\|B(k)\|/\P[\O_1]
\Eq(4.233)
$$
It is easy to show that (see [B]) that
$$
\E\|B(k)\|\leq c\left(1+\sqrt{M/k}\right)^2
\Eq(4.234)
$$
for some constant $2>c>1$.
Collecting our estimates and using that $1+x\leq e^x$ we arrive at
$$
\E\left[e^{u f_N^{(k)}}\big|\tilde \FF_{k+1}\right]
 \leq \exp\left(\frac 12 u^2 e^{|u|12\sqrt M/N}N^{-2}4\left[
8\a^2 +76 (1+\sqrt{M/k})^2\right]\right)
\Eq(4.235)
$$
Since
$$
\sum_{k=1}^N  (1+\sqrt{M/k})^2 = N + 4 \sqrt {MN}+ M\ln N
=N(1+4\sqrt\a +\a \ln N)
\Eq(4.236)
$$
this yields that
$$
\P\left[\sum_{k=1}^N f_N^{(k)}\geq x\big|\O_1\right]
\leq \inf_u \exp\left(-ux + \frac { u^2}{2N} e^{|u|12\sqrt M/N}4
\left[ 8 \a^2 +76 +304 \sqrt\a +76 \a \ln N\right]\right)
\Eq(4.237)
$$
In order to perform the infimum over $u$ in \eqv(4.237) we must
distinguish two cases. First, if $\a\leq1/\ln N$, we may chose
$u=\sqrt{N}$ which yields
$$
\P\left[\sum_{k=1}^N f_N^{(k)}\geq x\right]
\leq e^{-\sqrt{N}x+c_1}
\Eq(4.237bis)
$$
for some positive constant $c_1$. If now $\a$ goes to zero with $N$ more
slowly than $1/\ln N$, a good estimate of the
infimum is obtained by choosing $u=N/12\sqrt M$.
This gives
$$
\eqalign{
\P\left[\sum_{k=1}^N f_N^{(k)}\geq x\right]
&\leq e^{-\sqrt N\frac {x}{12\sqrt\a}} \exp\left\{\frac {e}{9}
\left[\a+\frac{12}{\a}+\frac{48}{\sqrt\a}+2\ln N\right]\right\}\cr
&\leq e^{-\sqrt N x/12 +c_2\ln N}\cr
}
\Eq(4.238)
$$
for some positive constant $c_2$.
{}From here the lemma follows immediately. \endproof

\corollary {4.5} {\it There exists a set $\O_3(t^*)\subset \O_1$ with
$\P[\O_1\ba\O_3]\leq e^{-b N^{1/4} }$ such that for all
$\o\in \O_3(t^*)$
$$
\tilde\P_\s\left[\left\|m_N(\s)-m^*\right\|_2\leq [2(2\a + N^{-1/4})]^{1/2}
\right]\geq \frac 12
\Eq(4.239)
$$
}

\proof This follows from combining \eqv(4.11) and \eqv(4.13) with
Lemmas 4.2, 4.3 and 4.4 and choosing  $x=N^{-1/4}$ in the latter.\endproof

Since by assumption $\|t^*\|_2<c$, lemma 2.3 implies that on
$\O_1$, $\|m^*\|_2\leq 2$.
As a consequence, putting together Proposition 3.1, Corollary 4.5 and
\eqv(2.10), we find that on $\O_3(t^*)$,
$$
\frac 1{\b N}\ln Z_{N,\b,\e+\rho}^I[\o](\tilde m)\geq
\E \Psi_{N}(m^*,t^*)-10\a^{1/2-\d} -\rho (c+2-\rho/2)-\frac {\ln 2}{\b N}
\Eq(4.240)
$$
Which is the desired form of the lower bound.

Finally, by a simple Borel-Cantelli argument, it follows from the
estimates on
the probabilities of the sets $\O_1,\O_2$ and $\O_3(t^*)$ that
there exits a set $\tilde \O$ of measure one on which
$$
\limsup_{N\uparrow\infty}\frac 1{\b N}\ln Z_{N,\b,\e}^I[\o](\tilde m)
\leq \limsup_{N\uparrow\infty}
\sup_{m:\,\|\Pi_I m-\tilde m\|_2\leq \e}\inf_{t\in\R^M}\E \Psi_{N}(m,t)
\Eq(4.23)
$$
and
$$
\liminf_{N\uparrow\infty}\frac 1{\b N}\ln Z_{N,\b,\e}^I[\o](\tilde m)
\geq \liminf_{N\uparrow\infty}
\sup_{m:\,\|\Pi_I m-\tilde m\|_2\leq \e-\rho}\inf_{t\in\R^M}\E \Psi_{N}(m,t)
\Eq(4.24)
$$
It remains to show that the limsup and the liminf's on the right-hand sides
of \eqv(4.23) and \eqv(4.24) coincide.
{}From here on there is no difference to the procedure in the case
$M<\ln N/\ln 2$ that was treated in [BGP2]. We repeat the outline for
the convenience of the reader. We write $\E\Psi_N(m,t)$ in its
explicit form as
$$
\E\Psi_N(m,t)=\frac 12\|m\|_2^2-(m,t)+\b^{-1}
2^{-M}\sum_{\g=1}^{2^M} \ln\cosh(\b(e_\g,t))
\Eq(4.25)
$$
where the vectors $e_\g$, $\g=1,\dots,2^M$ form a complete
enumeration of all vectors with components $\pm 1$ in $\R^M$.
They can be conveniently chosen as
$$
e_\g^\mu=(-1)^{\left[\g 2^{1-\mu}\right]}
\Eq(4.26)
$$
where $[x]$ denotes the smaller integer greater or equal to $x$.
Note that $\E\Psi_N(m,t)$ depends on $N$ only through $M(N)$. We
may use Lemma 2.3 to show that
$$
\inf_{t\in\R^M}\E \Psi_{N}(m,t)=\frac 12\|m\|_2^2-
\inf_{y\in \R^ {2^M}:\, n_{p}(y)=m}
\b^{-1}2^{-M}\sum_{\g=1}^{2^M} I(y_\g)
\Eq(4.27)
$$
and hence
$$
\sup_{m:\,\|\Pi_I m-\tilde m\|_2\leq \e}\inf_{t\in\R^M}\E \Psi_{N}(m,t)
=\sup_{y\in \R^ {2^M}:\, \left\|\Pi_I n_{p}(y)-\tilde m\right\|_2\leq \e}
\frac 12\left\|n_{p}(y)\right\|_2^2-\b^{-1}2^{-M}\sum_{\g=1}^{2^M} I(y_\g)
\Eq(4.28)
$$
To prove that this expression converges as $N$ (or rather $M$)
tends to infinity,
we define  the sets
$$
\AA_d^M\equiv \left\{ y\in [-1,1]^{2^M}\,\big | \,
y_\g=y_{\g+2^d}\right\}
\tag {4.29}
$$
Obviously,
$$
\AA^M_0\subset\AA^M_1\subset\dots \AA^M_{M-1}\subset
\AA_{M}^M=[-1,1]^{2^M}
\tag {4.30}
$$
The definition of these sets implies the following fact: If $y\in
\AA^M_d$ with $d<M$, then
\item{(i)} $n^\nu_{p}(y)=0$, if $\nu>d$ and
\item{(ii)} $n^\mu_{p}(y)=n_d^\mu(y)$, if $\mu\leq d$.

Let us set
$$
\Theta_{p}(y)=  \frac 12\left\|n_{p}(y)\right\|_2^2-\b^{-1} 2^{-p}
\sum_{\g=1}^{2^p}
I(y_\g)
\tag {4.32}
$$
and
$$
\Upsilon_{p,\e}(\tilde m)=\sup_{{y\in \AA^M_{p}}\atop
{\|\Pi_I n_{p}(y)-\tilde m\|_2\leq \e}}
               \Theta_{p}(y)
\tag{4.33}
$$
Therefore, for $y\in\AA_d^p$, $\Theta_{p}(y)=\Theta_d(y)$,
while at the same time the constraint in the sup is satisfied
simultaneously w.r.t. $n_{p}$ or $n_d$, as soon as $d$ is large
enough  such that $I\subset
\{1,\dots,d\}$. Therefore,
$$
\Upsilon_{p,\e}(\tilde m) \geq \sup_{{y\in \AA^M_d}\atop
{\|\Pi_I n_{p}(y)-\tilde m\|_2\leq \e}}
\Theta_{p}(y)= \sup_{{y\in \AA^d_d}\atop
{\|\Pi_I n_{p}(y)-\tilde m\|_2\leq \e}}
\Theta_d(y)=\Upsilon_{d,\e}(\tilde m)
\tag{4.34}
$$
Hence $\Upsilon_{p,\e}(\tilde m)$ is an increasing sequence in $M$ and
being bounded from above, converges.
Thus
$$
\eqalign{
\lim_{N\uparrow\infty}
\sup_{m:\,\|\Pi_I m-\tilde m\|_2\leq \e}\inf_{t\in\R^M}\E \Psi_{N}(m,t)&=
\lim_{N\uparrow\infty} \Upsilon_{p,\e}(\tilde m)\cr
&=\sup_{p}\Upsilon_{p,\e}(\tilde m)
}
\Eq(4.35)
$$
It remains to consider the limit $\e\downarrow 0$. It is clear
that
 $\sup_{p} \Upsilon_{p,\e}(\tilde m)$ converges to a
 lower-semicontinuous function and that
 $$
 \lim_{\e\downarrow 0} \sup_{p} \Upsilon_{p,\e}(\tilde m)
 =  \lim_{\e\downarrow 0}\sup_{m:\,\|\Pi_I m-\tilde m\|_2\leq \e}
 \sup_{p} \Upsilon_{p,0}(\tilde m)
\Eq(4.36)
$$
Thus if $ \sup_{p} \Upsilon_{p,0}(\tilde m)$ is continuous in a
neighborhood of $\tilde m$, we get
$$
 \lim_{\e\downarrow 0} \sup_{p} \Upsilon_{p,\e}(\tilde m)
= \sup_{p} \Upsilon_{p,0}(\tilde m)
\Eq(4.37)
$$
as desired. But, as has been shown in [BGP2], from the explicit
form of $\Upsilon$ one shows easily that
$ \sup_{p} \Upsilon_{p,0}(\tilde m)$ is Lipshitz continuous in the
interior
of the set on which it is bounded. This proves Theorem 4.1
\endproof

We will show next that a sufficient condition for condition \eqv(4.1) to
hold is that $\tilde m$ belongs to $D_{|I|}$.
While this appears intuitively `clear', the rigorous proof is surprisingly
tedious.
 Let us first
introduce some notation and results.

Let $E_{p}$ be the $p\times 2^p$-matrix whose rows are given
by the vectors $e_{\g}$, $\g=1,\dots,2^p$, which, for convenience,
are ordered accordingly to \eqv(4.26). We will denote by $e^{\mu}$,
$\mu=1,\dots,p$ the column vectors of $E_{p}$ and by $E^t_{p}$ its
transpose. It can easily be verified that
$$
2^{-p}(e^{\mu},e^{\nu})=\cases{1 & if $\mu=\nu$\cr
                               0 & otherwise}
\Eq(4.38)
$$
Thus, the $2^p\times 2^p$-matrix
$2^{-p}E_{p}E^t_{p}$ is a projector that projects on the subspace spanned
by the   orthogonal vectors $\{e^{\mu}\}_{\mu=1}^p$, and
$2^{-p}E^t_{p}E_{p}$ is the identity in $\R^p$.
Given a linear transformation $A$ from $R^p$ to $R^q$, we define
$$
AC=\left\{Ax \mid  x\in C \right\} \,\,\,\,\hbox{for}\,\,\,\,
C\subset \R^p
\Eq(4.39)
$$
With this notations the vector $n_{p}(y)$ and the set $D_{p}$, defined in
\eqv(1.61) and \eqv(1.62), can be rewritten as
$$
\eqalign{
n_{p}(y)&=2^{-p}E^t_{p}y\cr
D_{p}&=2^{-p}E^t_{p}[-1,1]^{2^p}\cr
}
\Eq(4.40)
$$
Moreover, for any set $I\subset\{1,\dots, p\}$, we have the following
property,
$$
\Pi_ID_{p}=D_{|I|}
\Eq(4.41)
$$

Finally, let us remark that of course the statements of Lemma 2.3
apply also to the  deterministic
function $\inf_{t\in \R^p} \E \Psi_{N,\b}(m,t)$ . All references to
Lemma 2.3 in the sequel are to be understood as referring to properties of
this latter  function.

By Lemma 2.3, the
condition \eqv(4.1) of Theorem 4.1 is satisfied if and only if
the supremum in the l.h.s of \eqv(4.1) is taken on at a point
$m$ in $\hbox{int}D_{p}$.
More precisely, by  \eqv(2.anton),
this condition is equivalent to demanding that
for all $\e>0$ and all $p$, the supremum over $y$ s.t.
$\|\Pi_In_p(y)-\tilde m\|_2\leq \e$ of $\Theta_p(y)$
is taken on at a point $y^*$ such that
$$
2^{-p}\sum_{\g=1}^{2^p} \left[I'(y_\g^*)\right]^2 \leq c
\Eq(t1)
$$
We set
$$
{\cal A}_\e(\tilde m)\equiv
\left\{y\in[-1,1]^{2^M}:\,\|\Pi_I n_{p}(y)-\tilde m\|_2\leq \e\right\}
\Eq(4.47)
$$
\def\mod{\hbox{\ftn mod}}
\def\gt{{\tilde\g}}
\def\Th{\Theta}

\lemma {4.6} {\it Assume that $0<\b<\infty$. Then for all $\tilde m\in D_{|I|}$
and
$\e>0$ there exists $c(\tilde m,\e)<\infty$
such that for all $p\geq |I|$
$$
\inf_{{y\in [-1,1]^{2^p}}\atop{\Pi_I n_p(y)\in B_\e(\tilde m)}}
\Th (y)=\Th(y^*)
\Eq(p1)
$$
where
$$
T_p(y^*)\equiv
2^{-p}\sum_{\g=1}^{2^p}\left[I'(y_\g^*)\right]^2\leq c(\tilde m,\e)
\Eq(p2)
$$
}

\proof  The proof proceeds by showing that if $y$ does not
satisfy condition \eqv(p2), then we can find a $\d y$ such that
$y+\d y\in \A_\e(\tilde m) $ and $\Th_p(y+\d y)<\Th_p(y)$, so that
$y$ cannot be the desired $y^*$. Let us first note that
$$
\Th_{p}(y+\d y)-\Th_{p}(y)
=\frac{1}{2}\left[\|n_{p}(y+\d y)\|_2^2-\|n_{p}(y)\|_2^2\right]
+2^{-p}\b^{-1}\sum_{\g=1}^{2^{p}}
\left[I(y_{\g})-I(y_{\g}+\d y_{\g})\right]
\Eq(4.49)
$$
Using the properties of the matrix $E_{p}$ we can bound the difference of
the quadratic terms as follows
$$
\eqalign{
\|n_{p}(y+\d y)\|_2^2-\|n_{p}(y)\|_2^2
=&\|n_{p}(\d y)\|_2^2+2^{-p+1}(\d y, 2^{-p}E_{p}E^T_{p} y)\cr
\geq& -2^{-p/2+1}\|\d y\|_2\cr
}
\Eq(4.50)
$$
Thus we can show that $\Th_p(y+\d y)<\Th_p(y)$ holds by showing
that
$$
2^{-p}\b^{-1}\sum_{\g=1}^{2^{p}}
\left[I(y_{\g})-I(y_{\g}+\d y_{\g})\right]
>2^{-p/2}\|\d y\|_2
\Eq(4.51)
$$
Define the map $Y$  from $[-1,1]^{2^p}$ to $[-1,1]^{2^{|I|}}$
by
$$
Y_{\g}(y)\equiv 2^{-p+|I|}\sum_{\tilde\g=0}^{2^{p-|I|}-1}
y_{\g+\tilde\g 2^{|I|}}\,\,\,\,\,,\,\,\,\, \g=1,\dots,2^{|I|}
\Eq(4.53)
$$
 Using \eqv(4.40) we get that
$$
\eqalign{
\Pi_{|I|}n_{p}(y)
&=2^{-|I|}E^t_{|I|}\left(2^{-p+|I|}\sum_{\g=1}^{2^{p-|I|}}
\Pi_{\{(\g-1)2^{|I|}+1,\dots,\g2^{|I|}\}}y\right)
=2^{-|I|}E^t_{|I|}Y(y)
}
\Eq(4.52)
$$
Therefore, the property that  $y\in \A_\e(\tilde m)$  depends
only on the quantity $Y(y)$.

Notice that if $\tilde m\in D_{|I|}$ and $\e>0$, then there exists
$X\in (-1,1)^{2^{|I|}}$ such that $\|n_I(X)-\tilde m\|_2\leq \e$.
This implies that for any $p$, the vector $x\in \R^{2^p}$ with components
$x_\g\equiv X_{y\,\mod\, 2^{|I|}}$ lies also in $\A_\e(\tilde m)$.
Moreover,
$$
\max_{\g} |x_\g|=\max_{\g}|X_\g|\equiv 1-d<1
\Eq(p4)
$$
and therefore
$$
T_p(x)\leq  \left[I'(1-d)\right]^2
\Eq(p5)
$$
is some finite $p$-independent constant. We will use this fact to
construct our $\d y$. We may of course choose an optimal $X$,
i.e. one for which $d$ is maximal. In the sequel $X$ and $x$ will
refer to this vector. Let now $y$ be a vector in $\A_\e(\tilde
m)$ for which $T_p(y)> c$ for some large constant $c$. We will
show that this cannot minimize $\Th_p$. We will distinguish two
cases:

\noindent {\bf Case 1: } Let us introduce two parameters, $0<\eta\ll d$ and
$0<\lambda<1$, that will be appropriately chosen later. In this first case we
assume that $y$ is such that
$$
\sum_{\g=1}^{2^p}\1_{\left\{|y_{\g}|\geq 1-\eta\right\}}
\geq (1-\lambda)2^{p-|I|}
\Eq(4.55ter)
$$
and we choose
$$
\d y\equiv \rho (x-y)
\Eq(4.56)
$$
where $0<\rho<1$ will be determined later.
It then trivially follows from the definition of $x$ and the
convexity of the set $\A_\e$ that
$y+\rho (x-y)\in\A_\e$ and that $y+\rho(x-y)\in [-1+\rho d,1-\rho
d]^{2^p}$. If
Thus if we can  show that with this choice, and with an $\rho$ such
that $\rho d>\eta$,  \eqv(4.51) holds,
we can exclude that the infimum is  taken on for such an $y$.

Let us first consider components $y_{\g}$ such
that $|y_{\g}|>1-d$. Since $|x_{\g}|\leq 1-d$ we have, for those
components, $\hbox{sign}\d y_{\g}=-\hbox{sign}y_{\g}$ and thus
$I(y_{\g})-I((y+\d y)_{\g})>0$. This fact together with \eqv(4.55ter)
entails
$$
\eqalign{
2^{-M}\sum_{\g=1}^{2^p}[I(y_{\g})-I((y+\d y)_{\g})]
 \1_{\left\{|y_{\g}|\geq 1-d\right\}}
\geq& 2^{-M}\sum_{\g=1}^{2^p}[I(y_{\g})-I((y+\d y)_{\g})]
\1_{\left\{|y_{\g}|\geq 1-\eta\right\}}\cr
\geq &\inf_{{|y_{\g}|\geq 1-d}\atop{|x_{\g}|\leq 1-d}}
(1-\lambda)2^{-|I|}[I(y_{\g})-I((y+\d y)_{\g})]\cr
}
\Eq(4.58bis)
$$
Note that $I(z)$ is symmetric with respect to zero and is a strictly
increasing function of $z$ for $z>0$. Thus $I((y+\d y)_{\g})$ is maximized
over $|x_{\g}|\leq 1-d$ for $x_{\g}=(1-d)\hbox{sign}y_{\g}$. From
this we get
$$
\inf_{|x_{\g}|\leq 1-d}[I(y_{\g})-I((y+\d y)_{\g})]
\geq [I(y_{\g})-I(|y_{\g}|+\rho((1-d)-|y_{\g}|))]
\Eq(4.58ter)
$$
and the infimum over $|y_{\g}|\geq 1-\eta$ in the r.h.s. of \eqv(4.58ter)
is easily seen to be taken on for $|y_{\g}|= 1-\eta$. Thus
$$
\eqalign{
\inf_{{|y_{\g}|\geq 1-d}\atop{|x_{\g}|\leq 1-d}}
(1-\lambda)2^{-|I|}[I(y_{\g})-I((y+\d y)_{\g})]
\geq &(1-\lambda)2^{-|I|}[I(1-\eta)-I(1-\eta-\rho(d-\eta))]\cr
\geq & (1-\lambda)2^{-|I|} \rho(d-\eta)I'(1-\eta-\rho(d-\eta))\cr
\geq & (1-\lambda)2^{-|I|} \rho(d-\eta)\frac{1}{2}|\ln(\eta+\rho(d-\eta))|
}
\Eq(4.59)
$$
where we have used the convexity of $I$ and the bound,
$
I'(1-x)\geq\frac{1}{2}|\ln x|
$
for $0<x<1$.

We now have to consider the components $y_{\g}$ with $|y_{\g}|\leq 1-d$.
Here the entropy difference $I(y_{\g})-I((y+\d y)_{\g})$ can of course be
negative. To get a lower bound on this difference we use \eqv(4.58ter) and
perform the change of variable $|y_{\g}|=(1-d)-z_{\g}$ to write
$$
\eqalign{
\inf_{{|y_{\g}|\leq 1-d}\atop{|x_{\g}|\leq 1-d}}
[I(y_{\g})-I((y+\d y)_{\g})]
=&\inf_{0\leq z_{\g}\leq 1-d}
I((1-d)-z_{\g}+\rho z_{\g})-I((1-d)-z_{\g})\cr
\geq& \inf_{0\leq z_{\g}\leq 1-d}-\rho z_{\g} I'((1-d)-z_{\g}+\rho z_{\g})\cr
=&\inf_{0\leq z_{\g}\leq 1-d}-\rho z_{\g}\frac{1}{2}\ln\left(
\frac{2-d-z_{\g}+\rho z_{\g}}{d+z_{\g}-\rho z_{\g}}\right)\cr
\geq &-\rho (1-d)\frac{1}{2}\ln\left(\frac{2-d}{d}\right)\cr
\geq &-\frac{\rho}{2}\ln \frac{2}{d}\cr
}
\Eq(4.61)
$$
and putting together \eqv(4.61) and \eqv(4.55ter) yields
$$
2^{-M}\sum_{\g=1}^{2^p}[I(y_{\g})-I((y+\d y)_{\g})]
\1_{\left\{|y_{\g}|< 1-d\right\}}
\geq  -(1-(1-\lambda)2^{-|I|})
\frac{\rho}{2}\ln\left(\frac{2}{d}\right)
\Eq(4.63)
$$
Therefore, \eqv(4.63) together with \eqv(4.58bis) and \eqv(4.59) give
$$
\eqalign{
&\b^{-1}2^{-M}\sum_{\g=1}^{2^p}[I(y_{\g})-I((y+\d y)_{\g})]\cr
\geq &\b^{-1}\rho\left\{
(1-\lambda)2^{-|I|}(d-\eta)\frac{1}{2}|\ln(\eta+\rho(d-\eta))|
- (1-(1-\lambda)2^{-|I|})\frac{1}{2}\ln\left(\frac{2}{d}\right)
\right\}\cr
}
\Eq(4.64)
$$
On the other hand, we have
$$
2^{-M/2}\|\d y\|_2\leq 2\rho
\Eq(4.65)
$$
Consequently, \eqv(4.51) holds if we can choose $\lambda$, $\eta$, and
$\rho$ so that the following inequality holds,
$$
\b^{-1}\left\{
(1-\lambda)2^{-|I|}(d-\eta)\frac{1}{2}|\ln(\eta+\rho(d-\eta))|
- (1-(1-\lambda)2^{-|I|})\frac{1}{2}\ln\left(\frac{2 }{d}\right)
\right\}>2
\Eq(4.66)
$$
But this is always possible by taking e.g. $\lambda<1$,
$\eta\equiv\rho d/2$
and $\rho\equiv d^K$ where $K\equiv K(d, |I|, \lambda)>1$ is chosen
sufficiently large as to satisfy
$$
(1-\lambda)2^{-|I|}d(1-d^K/2)\frac{K+1}{2}|\ln d|
>4+\frac{1}{2}|\ln d|
\Eq(4.66bis)
$$

\noindent {\bf Case 2: } We will assume that $\lambda<1$, and that
$\eta$, and $\rho$ are chosen as in the  case 1.
We can then assume that
$$
\sum_{\g=1}^{2^p}\1_{\left\{|y_{\g}|\geq 1-\eta\right\}}< (1-\lambda)2^{p-|I|}
\Eq(4.56bis)
$$
We assume further that
$$
T_p(y)>c
\Eq(ppp)
$$ for $c$ sufficiently large to be chosen later.
Here we will choose $\d y$ such that
$$
Y(\d y)\equiv 0
\Eq(4.57)
$$
so that trivially $y+\d y\in \A_\e(\tilde m)$.
Let us introduce a parameter $0<\zeta<\eta$, that
we will choose appropriately later, and let us set,
for $\g\in\{1,\dots,2^{|I|}\}$,
$$
{\cal K}_{\g}^+
\equiv\{\tilde \g\in\{1,\dots, 2^{p-|I|}\}\bigm|
|y_{\g+(\tilde\g-1)2^{|I|}}|\geq 1-\zeta\}
\Eq(p6)
$$
and
$$
{\cal K}_{\g}^-
\equiv\{\tilde \g\in\{1,\dots, 2^{p-|I|}\}\bigm|
|y_{\g+(\tilde\g-1)2^{|I|}}|\leq 1-\eta\}
$$
For all indices $\g$ such that ${\cal K}_{\g}^+=\emptyset$, we simply set
$\d y_{\g+(\tilde\g-1)2^{|I|}}\equiv 0$ for all
$\tilde \g\in\{1,\dots, 2^{p-|I|}\}$.
If $\KK_\g^+$ were empty for {\it all} $\g$, then $T_p(y)\leq
[I'(1-\zeta)]^2$ which contradicts our assumption \eqv(ppp), for
suitably large $c$ (depending only on $\zeta$).
Thus we consider  now the remaining
indices $\g$ for which ${\cal K}_{\g}^+\neq\emptyset$.

 First note that \eqv(4.56bis) implies that
$|{\cal K}_{\g}^+|<(1-\lambda)2^{p-|I|}$ and that
${\cal K}_{\g}^->\lambda 2^{p-|I|}$ so that choosing
$1>\lambda >\frac{1}{2}$, we have $|{\cal K}_{\g}^+|<|{\cal K}_{\g}^-|$.
Our strategy will be to find $\d y$ in such a way as to
decrease the moduli of the components in $\KK_\g^+$ at the expense
of possibly increasing them on $\KK_\g^-$ in such a way as to
leave $Y(y+\d y)=Y(y)$.

We will in the sequel consider the case where there is only one
index $\g$, e.g. $\g=1$,  for which  $\KK_\g^+$ is nonempty.
The general case is treated essentially by iterating the same
procedure.
We will use the
simplified notation
$y_{1+2^{|I|}\gt}\equiv y_{\gt}$,
$\d y_{1+2^{|I|}\gt}\equiv \d y_{\gt}$
and also set $\KK_1^\pm\equiv
\KK^\pm$. We will  assume moreover that
all components  $y_\gt$ are positive, as this is
the worst situation. We will chose $\d y$ such that $\d y_\gt=0$, if
$\gt\in \{\KK^+\cup\KK^-\}^c$ and $\d y_\gt<0 $ if $\gt\in \KK^+$.
For each $\gt\in\KK^+$ we will choose in a unique and distinct
 $\gt'\in  \KK^-$ and set $\d y_{\gt'}=-\d y_\gt$. This ensures
 that $Y(\d y)=0$. We will also make sure that for all $\gt$,
 $|\d y\gt|\leq \eta/2-\zeta$.

We have to construct $\d y_\gt$ for $\gt\in \KK^+$. In this
process we have to consider the following three functionals:
\item{(1)} The change in the quadratic term of $\Th_p$. This is
bounded  by
$$
\d E(\d y) \equiv 2^{-p/2+1} \sqrt{2\sum_{\gt\in \KK^+} \d y_\gt^2}
\Eq(p8)
$$
\item {(2)} The change in the entropy term,
$$
\eqalign{
\d I(\d y)&=2^{-p} \sum_{\gt\in \KK^+} \left(I(y_\gt+\d y_\gt)
-I(y_\gt)\right)\cr
&+2^{-p} \sum_{\gt\in \KK^-} \left(I(y_\gt+\d y_\gt)-I(y_\gt)
\right)    \cr
&\geq
2^{-p} \sum_{\gt\in \KK^+} |\d y_{\gt}| I'(y_\gt+\d y_\gt)
-2^{-p} \sum_{\gt\in \KK^-} \d y_{\gt}|\ln\eta|/2\cr
&=2^{-p} \sum_{\gt\in \KK^+} |\d y_{\gt}|\left(I'(y_\gt+\d
y_\gt)-|\ln \eta|/2\right)
\cr
&\geq  2^{-p-1} \sum_{\gt\in \KK^+} |\d y_{\gt}|I'(y_\gt+\d
y_\gt)
}
\Eq(p9)
$$
where we have used that for $1\geq |x|\gg |y|\gg 0.9$,
$I(x)-I(y)\approx |x-y| |\ln (1-y)|$ and that
under our assumption for $\gt\in \KK^+$, $y_\gt+\d y_\gt\geq
1-\eta/2$.
\item{(3)} Finally, we have that
$$
\eqalign{
T_p(y+\d y)&\leq  2^{-p} \sum_{\g\not\in\KK^+}
[I'(y_\gt+\d y_\gt)]^2 +2^{-p} \sum_{\gt\in\KK^+}
[I'(y_\gt+\d y_\gt)]^2\cr
&\leq [I'(1-\eta/2)]^2 + 2^{-p} \sum_{\gt\in\KK^+}
[I'(y_\gt+\d y_\gt)]^2
}
\Eq(p10)
$$

Looking at these three functionals suggests to choose
$\d y_\gt$ for $\gt \in \KK^+$  as the solution of the equation
$$
-\d y_\gt = \t I'(y_\gt+\d y_\gt)
\Eq(p11)
$$
The point is that with this choice \eqv(p10) yields
(we set for simplicity $\d E(\d y(\t))\equiv\d E(\t)$, etc.)
$$
\d I(\t) \geq \frac {1}{8 \t} (\d E(\t))^2
\Eq(p12)
$$
while
$$
T_p(\t)\leq [I'(1-\zeta)]^2+\t^{-2} (\d E(\t))^2
\Eq(p13)
$$
Thus we can ensure that the entropy gain dominates the potential
loss in the quadratic term provided we can choose
$\t< \d E(\t)/8$.
However, we know that $T_p(\t)$ is a continuous function
of $t$ and $T_p(0) \geq c$. Thus there exists $\t_0>0$ such that
for all $\t\leq\t_0$, $T_p(\t)\geq c/2$, and so by
\eqv(p13),
$$
\t^{-1} \d E(\t)\geq \sqrt{c/2- [I'(1-\zeta)]^2}
\Eq(p14)
$$
which inserted in \eqv(p13) yields that
$$
\d I(\t) \geq \frac {\ln 2}{4} \sqrt{c/2- [I'(1-\zeta)]^2} \d
E(\t)
\Eq(p15)
$$
It is clear that if $c$ is chosen large enough (`large' depending
only on $\zeta$), this gives $\d I(t)>\d E(t)$, as desired.
Finally, it is easy to see that $|\d y_\gt|$ is bounded from
above by the solution of the equation
$$
x=\t I'(1-x)
\Eq(p16)
$$
which is of the order of $x\approx \t |\ln\t|$. If $\zeta$ is
chosen e.g. $\zeta=\eta/4$, we see from this that
for small enough $\t$, $|\d y_\gt|\leq \eta/2-\zeta$, so that all
our conditions can be satisfied. Thus, there exist $c<\infty$ depending
only on $\eta$ (which in turn depends only on $\tilde m$ and $\e$)
such that any $y$ that satisfies the assumptions of Case 2 with
this choice of $c$ in \eqv(ppp) cannot realize the infimum of
$\Th_p$.
The two cases combined prove the lemma.
\endproof

To conclude the proof of Theorem 1 we show that for $\tilde m\in D_I^c$
\eqv(1.91) holds. This turns out to be rather simple.
 The main idea is that if $\tilde m\in D_{|I|}^c$, then
on a subset of $\O$ of probability one, for $N$ large enough and $\e$
small enough, the set
$\left\{\s\in{\cal S}_N\mid \|\Pi_{I}m_N(\s)-\tilde m\|_2\leq\e\right\}$
is empty.

To do so we will first show that uniformly in the configurations $\s$,
the vector $\Pi_{I}m_N(\s)$ can be rewritten as the sum of a vector
in $D_{|I|}$ and a vector whose norm goes to zero as $N$ goes to infinity.
Let $e_{\g}$, $\g=1,\dots,2^{|I|}$, be the column vectors of the matrix
$E^t_{|I|}$. We set
$$
v_{\g}\equiv\left\{ i\in\{1,\dots,N\} \mid \xi^{\mu}_i=e^{\mu}_{\g}\,\,\,,
\,\,\,\forall \mu\in I\right\}
\Eq(4.78)
$$
These sets are random sets, depending on the realization of the random
variables $\xi^{\mu}_i$. Their cardinality, however, remains very close
to their mean value. More precisely let $\l_{\g}$ denote the fluctuation
of $|v_{\g}|$ about its mean,
$$
\l_{\g}\equiv 2^{|I|}N^{-1}\left||v_{\g}|-2^{-|I|}N\right|
\Eq(4.79)
$$
There exists a subset $\O_4\in\O$ of
probability one and a function, $\d_N$, tending to zero as $N$ tends to
infinity, such that for all but a finite number of indices,
$$
|\l_{\g}|<\d_N\,\,\,\,,\,\,\,\,\g=1,\dots,2^{|I|}
\Eq(4.80)
$$
This fact has been proven in [G].  Using \eqv(4.78),
$\Pi_{I}m_N(\s)$ can be rewritten as
$$
\Pi_{I}m_N(\s)=2^{-|I|}E^t_{|I|}(X(\s)+\d X(\s))
\Eq(4.81)
$$
where $X(\s)$ and $(\d X)(\s)$ are respectively the vectors with
components
$X_{\g}(\s)\equiv |v_{\g}|^{-1}\sum_{i\in v_{\g}}\s_i\in [-1,1]$,
$(\d X)_{\g}(\s)\equiv\l_{\g}X_{\g}(\s)$, $\g=1,\dots,2^{|I|}$.
It then follows from the
properties of the matrix $E^t_{|I|}$ and \eqv(4.80) that, on $\O_4$,
$$
\left\|
\Pi_{I}m_N(\s)-n_{|I|}(X(\s))
\right\|_2<\d_N
\Eq(4.811)
$$

Now, by assumption, $\tilde m\in D_{|I|}^c$, i.e.
there exists $\tilde\e>0$ such that
$\{x\in\R^{|I|}\mid\|x-\tilde m\|_2\leq\tilde\e\}
\subset\left(D_{|I|}\right)^c$. Therefore, since
$n_{|I|}(X(\s))\in D_{|I|}$, we have
$\|n_{|I|}(X(\s))-\tilde m\|_2>\tilde\e$. From this and \eqv(4.811) it
follows that on $\O_4$,
$\|\Pi_{I}m_N(\s)-\tilde m\|_2>\tilde\e-\d_N$. Finally, for $N$ large
enough and $\e$ small enough we get
$$
\left\{\s\in{\cal S}_N\mid \|\Pi_{I}m_N(\s)-\tilde m\|_2\leq\e\right\}
=\emptyset
\Eq(4.82)
$$
{}From this, part \eqv(1.91) easily follows.
This concludes the proof of Theorem 1.

\newpage

\frenchspacing
\chap{References}4
\item{[AGS]} D.J. Amit, H. Gutfreund and H.
Sompolinsky, ``Statistical mechanics of neural networks near
saturation'', Ann. Phys. {\bf 173}: 30-67 (1987).
\item{[B]} A. Bovier, ``Self-averaging in a class of generalized
Hopfield models'', J. Phys. A  {\bf 27}: 7069-7077 (1994).
\item{[BG1]} A. Bovier and V. Gayrard, ``Rigorous results on the
thermodynamics of the dilute Hopfield model'', J. Stat. Phys. {\bf 69}:
597-627 (1993).
\item{[BGP1]} A. Bovier, V. Gayrard, and P. Picco, ``Gibbs states
of the Hopfield model in the regime of perfect memory'',
 Prob. Theor. Rel. Fields {\bf 100}: 329-363 (1994).
\item{[BGP2]} A. Bovier, V. Gayrard, and P. Picco,
``Large deviation principles for the Hopfield model and the
Kac-Hopfield model'', to appear in Prob. Theor. Rel. Fields (1995).
\item{[BGP3]} A. Bovier, V. Gayrard, and P. Picco,
``Gibbs states of the Hopfield model with extensively many patterns'',
 J. Stat. Phys. {\bf 79}: 395-414 (1995).
\item{[Co]} F. Comets, ``Large deviation estimates for a conditional
probability distribution. Applications to random Gibbs measures'',
Prob. Theor. Rel. Fields {\bf 80}: 407-432 (1989).
\item{[DZ]} A. Dembo and O. Zeitouni, Large deviation techniques and
applications, Jones and Bartlett, Boston (1992).
\item{[E]} R.S. Ellis, ``Entropy, large deviations, and statistical
mechanics'',  Springer, Berlin (1985).
\item{[FP1]} L.A. Pastur and A.L. Figotin, ``Exactly soluble model
of a spin glass'', Sov. J. Low Temp. Phys. {\bf 3(6)}: 378-383
(1977).
\item{[FP2]} L.A. Pastur and A.L. Figotin, ``On the theory of
disordered spin systems'', Theor. Math. Phys. {\bf 35}: 403-414
(1978).
\item{[G]} V. Gayrard, ``Thermodynamic limit of the $q$-state
Potts-Hopfield model with infinitely many patterns'',
J. Stat. Phys. {\bf 68}: 977-1011 (1992).
\item{[Ho]} J.J. Hopfield,``Neural networks and physical systems
with emergent collective computational abilities'', Proc. Natl.
Acad. Sci. USA {\bf 79}: 2554-2558 (1982).
\item{[K]} H. Koch, ``A free energy bound for the Hopfield
model'', J. Phys. A {\bf 26}: L353-L355    (1993).
\item{[KP]} H. Koch and J. Piasko, ``Some rigorous results on the Hopfield
neural network model'', J. Stat. Phys. {\bf 55}: 903-928 (1989).
\item{[KPa]} J. Koml\'os and R. Paturi,
``Convergence results in a autoassociative memory model'',
Neural Networks {\bf 1}: 239-250 (1988).
\item{[LT]} M. Ledoux and M. Talagrand, ``Probability in Banach spaces'',
Springer, Berlin-Heidelberg-New York, (1991).
\item{[N]} Ch.M. Newman, ``Memory capacity in neural network
models: Rigorous lower bounds'', Neural Networks {\bf 1}: 223-238
(1988).
\item {[P]} G. Parisi, ``Complexity in biology: the point of
view of a physicist'', Rome-preprint (1995).
\item{[PST]} L. Pastur, M. Shcherbina, and B. Tirozzi, ``The replica
symmetric solution without the replica trick for the Hopfield model'',
J. Stat. Phys. {\bf 74}: 1161-1183 (1994).
\item{[Ro]}  R.T. Rockafellar, ``Convex Analysis'', Princeton University
Press, Princeton (1970).
\item{[ST]} M. Shcherbina and B. Tirozzi, ``The free energy for a class
of Hopfield models'', J. Stat. Phys. {\bf 72}: 113-125 (1992).
\item{[Va]} S.R.S. Varadhan, ``Large deviations and applications'',
{\it SIAM}, Philadelphia Pensylvania (1984).
\item{[Yu]} V.V. Yurinskii, ``Exponential inequalities for sums of
random vectors'', J. Multivariate Anal. {\bf 6}: 473-499 (1976).

\end